\documentclass[longauth]{aa}  

\usepackage{graphicx}
\usepackage{txfonts}
\usepackage{natbib}
\bibpunct{(}{)}{;}{a}{}{,} 
\defcitealias{leger2009}{L{\'e}ger, Rouan, Schneider et al. 2009}

\newcommand{\corot}{\textsl{CoRoT}}

\def\ms{\mbox{\,m\,s$^{-1}$}}                    
\def\m2s2{\mbox{\,m$^{2}$\,s$^{-2}$}}            
\def\kms{\mbox{\,km\,s$^{-1}$}}                  
\def\kgm3{\mbox{\,kg\,m$^{-3}$}}                 
\def\gcm3{\mbox{\,g\,cm$^{-3}$}}                 
\def\degr{\mbox{$^{\circ}$}}                     
\def\chisq{\mbox{$\chi^{2}$}}                    
\def\redchisq{\mbox{$\chi^{2}_{\mathrm{red}}$}}  
\def\vsini{\mbox{$v$\,sin\,$i$}}                 
\def\teff{\mbox{$T_{\mathrm{eff}}$}}             
\def\logg{\mbox{$\log\,g$}}                      
\def\Msun{\mbox{$\mathrm{M}_{\odot}$}}           
\def\Rsun{\mbox{$\mathrm{R}_{\odot}$}}           
\def\Mjup{\mbox{$\mathrm{M}_{\mathrm{Jup}}$}}    
\def\Rjup{\mbox{$\mathrm{R}_{\mathrm{Jup}}$}}    
\def\Rs{\mbox{$R_{\mathrm{s}}$}}                 
\def\Rp{\mbox{$R_{\mathrm{p}}$}}                 
\def\Ms{\mbox{$M_{\mathrm{s}}$}}                 
\def\Mp{\mbox{$M_{\mathrm{p}}$}}                 
\def\Is{\mbox{$i_{\mathrm{s}}$}}                 
\def\Ip{\mbox{$i_{\mathrm{p}}$}}                 
\def\Av{\mbox{$A_{\mathrm{v}}$}}                 
\def\Rv{\mbox{$R_{\mathrm{v}}$}}                 
\def\Rseq{\mbox{$R_\mathrm{s,eq}$}}              
\def\Rspo{\mbox{$R_\mathrm{s,pole}$}}            
\def\aRs{\mbox{$a/\Rs$}}                         
\def\aRseq{\mbox{$a/\Rseq$}}                     
\def\krprs{\mbox{$k = \Rp/\Rs$}}                 
\def\Omegas{\mbox{$\Omega_{\mathrm{s}}$}}        
\def\OmegaRot{\mbox{$\Omega_{\mathrm{rot}}$}}    
\def\Tstar{\mbox{$T_{\mathrm{s}}$}}              
\def\Tpole{\mbox{$T_{\mathrm{pole}}$}}           
\def\Tspot{\mbox{$T_{\mathrm{spot}}$}}           
\def\gpole{\mbox{$g_{\mathrm{pole}}$}}           
\def\Prot{\mbox{$P_{\mathrm{rot}}$}}             
\def\Porb{\mbox{$P_{\mathrm{orb}}$}}             
\def\wPorbProt{\mbox{$w = \Porb/\Prot $}}        
\def\bspot{\mbox{$b_{\mathrm{spot}}$}}           
\def\lspot{\mbox{$l_{\mathrm{spot}}$}}           
\def\dspot{\mbox{$d_{\mathrm{spot}}$}}           

\newcommand{\angstrom}{\textup{\AA}}

\renewcommand{\theenumi}{\Alph{enumi}}

\begin{document} 

\title{Transiting exoplanets from the CoRoT space mission.
  \thanks{The CoRoT space mission, launched on December 27th 2006, was
    developed and is operated by CNES, with the contribution of
    Austria, Belgium, Brazil, ESA (RSSD and Science Programme),
    Germany, and Spain. 
    Based on observations obtained with the Nordic Optical Telescope,
    operated on the island of La Palma jointly by Denmark, Finland,
    Iceland, Norway, and Sweden, in the Spanish Observatorio del Roque
    de los Muchachos of the Instituto de Astrofisica de Canarias, in
    time allocated by OPTICON and the Spanish Time Allocation
    Committee (CAT).
    The research leading to these results has received funding from
    the European Community's Seventh Framework Programme
    (FP7/2007-2013) under grant agreement number RG226604 (OPTICON). 
    This work makes use of observations from the LCOGT network.
  }
}

\subtitle{XXVII. CoRoT-28b, a planet orbiting an evolved star, and CoRoT-29b, a planet showing an asymmetric transit} 

\author{
J.~Cabrera\inst{\ref{DLR}} 
\and Sz.~Csizmadia\inst{\ref{DLR}} 
\and G.~Montagnier\inst{\ref{IAP}}
\and M.~Fridlund\inst{\ref{Leiden},\ref{Onsala},\ref{MPIA}}
\and M.~Ammler-von Eiff\inst{\ref{Tautenburg}}
\and S.~Chaintreuil\inst{\ref{LESIA}}
\and C.~Damiani\inst{\ref{LAM}}
\and M.~Deleuil\inst{\ref{LAM}} 
\and S.~Ferraz-Mello\inst{\ref{Brasil}} 
\and A.~Ferrigno\inst{\ref{LESIA}}
\and D.~Gandolfi\inst{\ref{Heidelberg}}
\and T.~Guillot\inst{\ref{OCA}} 
\and E.~W.~Guenther\inst{\ref{Tautenburg},\ref{OAW}} 
\and A.~Hatzes\inst{\ref{Tautenburg}} 
\and G.~H\'ebrard\inst{\ref{IAP}} 
\and P.~Klagyivik\inst{\ref{IAC},\ref{ULL}} 
\and H.~Parviainen\inst{\ref{Oxford}}
\and Th.~Pasternacki\inst{\ref{DLR}}
\and M.~P{\"a}tzold\inst{\ref{Koeln}} 
\and D.~Sebastian\inst{\ref{Tautenburg}} 
\and M.~Tadeu dos Santos\inst{\ref{Brasil}} 
\and G.~Wuchterl\inst{\ref{Tautenburg}} 
\and S.~Aigrain\inst{\ref{Oxford}} 
\and R.~Alonso\inst{\ref{IAC},\ref{ULL}} 
\and J.-M.~Almenara\inst{\ref{LAM}} 
\and J.D.~Armstrong\inst{\ref{Hawaii},\ref{LCOGT}}
\and M.~Auvergne\inst{\ref{LESIA}} 
\and A.~Baglin\inst{\ref{LESIA}}
\and P.~Barge\inst{\ref{LAM}} 
\and S.~C.~C.~Barros\inst{\ref{LAM}} 
\and A.~S.~Bonomo\inst{\ref{Torino}}  
\and P.~Bord{\'e}\inst{\ref{Bordeaux}} 
\and F.~Bouchy\inst{\ref{LAM}} 
\and S.~Carpano\inst{\ref{MPE}} 
\and C.~Chaffey\inst{\ref{Torrance}}
\and H.~J.~Deeg\inst{\ref{IAC},\ref{ULL}} 
\and R.~F.~D{\'i}az\inst{\ref{Geneve}}
\and R.~Dvorak\inst{\ref{Wien}} 
\and A.~Erikson\inst{\ref{DLR}}
\and S.~Grziwa\inst{\ref{Koeln}}
\and J.~Korth\inst{\ref{Koeln}}
\and H.~Lammer\inst{\ref{OAW}} 
\and C.~Lindsay\inst{\ref{Iolani}}
\and T.~Mazeh\inst{\ref{Tel Aviv}}
\and C.~Moutou\inst{\ref{LAM},\ref{CFHT}} 
\and A.~Ofir\inst{\ref{Goettingen}}
\and M.~Ollivier\inst{\ref{IAS}} 
\and E.~Pall{\'e}\inst{\ref{IAC},\ref{ULL}}
\and H.~Rauer\inst{\ref{DLR},\ref{ZAA}} 
\and D.~Rouan\inst{\ref{LESIA}}
\and B.~Samuel\inst{\ref{LESIA}}
\and A.~Santerne\inst{\ref{CAUP}} 
\and J.~Schneider\inst{\ref{LUTh}} 
}

\institute{
Institute of Planetary Research, German Aerospace Center, Rutherfordstrasse 2, D-12489 Berlin, Germany\label{DLR}
\and Institut d'Astrophysique de Paris, UMR 7095 CNRS, Universit{\'e} Pierre \& Marie Curie, 98bis boulevard Arago, F-75014 Paris, France\label{IAP}
\and Leiden Observatory, University of Leiden, PO Box 9513, 2300 RA, Leiden, The Netherlands\label{Leiden}
\and Department of Earth and Space Sciences, Chalmers University of Technology, Onsala Space Observatory, 439 92, Onsala, Sweden\label{Onsala}
\and Max Planck Institute for Astronomy Königstuhl 17, D-69117 Heidelberg, Germany\label{MPIA}
\and Th{\"u}ringer Landessternwarte, Sternwarte 5, D-07778 Tautenburg, Germany\label{Tautenburg}
\and LESIA, UMR 8109 CNRS, Observatoire de Paris, UPMC, Universit{\'e} Paris-Diderot, 5 place J. Janssen, F-92195 Meudon, France\label{LESIA}
\and Aix Marseille Universit{\'e}, CNRS, LAM (Laboratoire d'Astrophysique de Marseille) UMR 7326, F-13388 Marseille, France \label{LAM}
\and IAG-Universidade de Sao Paulo, Brasil\label{Brasil}
\and Zentrum f{\"u}r Astronomie, Fakult{\"a}t für Physik und Astronomie, M{\"o}nchhofstr. 12-14, D-69120 Heidelberg, Germany\label{Heidelberg}
\and Universit{\'e} de Nice-Sophia Antipolis, CNRS UMR 6202, Observatoire de la C{\^o}te d'Azur, BP 4229, F-06304 Nice Cedex 4, France\label{OCA}
\and {\"O}sterreichische Akademie der Wissenschaften, Institut f{\"u}r Weltraumforschung, IWF, Schmiedlstra{\ss}e 6, 8042 Graz, Austria\label{OAW}
\and Instituto de Astrof{\'i}sica de Canarias, E-38205 La Laguna, Tenerife, Spain\label{IAC}
\and Dept. Astrof{\'i}sica, Universidad de la Laguna, E-38206 La Laguna, Tenerife, Spain\label{ULL}
\and Sub-department of Astrophysics, Department of Physics, University of Oxford, Oxford, OX1 3RH, UK\label{Oxford}
\and Rheinisches Institut f{\"u}r Umweltforschung an der Universit{\"a}t zu K{\"o}ln, Aachener Strasse 209, D-50931, Germany\label{Koeln}
\and University of Hawai'i Institute for Astronomy, 34 Ohia Ku Street, Pukalani, HI, USA\label{Hawaii}
\and Las Cumbres Observatory Global Telescope Network, Inc. 6740 Cortona Drive Suite 102, Goleta, CA 93117, USA\label{LCOGT}
\and INAF – Osservatorio Astrofisico di Torino, via Osservatorio 20, I-10025 Pino Torinese, Italy\label{Torino}
\and LAB, UMR 5804, Univ. Bordeaux \& CNRS, F-33270, Floirac, France\label{Bordeaux}
\and Max-Planck-Institut f{\"u}r extraterrestrische Physik, Giessenbachstrasse 1, D-85748 Garching, Germany\label{MPE}
\and Torrance High School, 2200 W Carson St, Torrance, CA 90501, USA\label{Torrance}
\and Observatoire de l'Universit{\'e} de Gen{\`e}ve, 51 chemin des Maillettes, CH-1290 Sauverny, Switzerland\label{Geneve}
\and University of Vienna, Institute of Astronomy, T{\"u}rkenschanzstr. 17, A-1180 Vienna, Austria\label{Wien}
\and Iolani School, 563 Kamoku St., Honolulu, HI-96816, USA\label{Iolani}
\and School of Physics and Astronomy, Raymond and Beverly Sackler Faculty of Exact Sciences, Tel Aviv University, Tel Aviv, Israel\label{Tel Aviv}
\and CFHT Corporation, 65-1238 Mamalahoa Hwy, Kamuela, Hawaii 96743, USA\label{CFHT}
\and Institut f{\"u}r Astrophysik, Friedrich-Hund-Platz 1, D-37077 G{\"o}ttingen, Germany\label{Goettingen}
\and Institut d'Astrophysique Spatiale, Universit{\'e} Paris-Sud \& CNRS, F-91405 Orsay, France\label{IAS}
\and Center for Astronomy and Astrophysics, TU Berlin, Hardenbergstr. 36, D-10623 Berlin, Germany\label{ZAA}
\and Instituto de Astrof\'isica e Ci\^{e}ncias do Espa\c{c}o, Universidade do Porto, CAUP, Rua das Estrelas, PT4150-762 Porto, Portugal\label{CAUP} 
\and LUTH, Observatoire de Paris, UMR 8102 CNRS, Universit{\'e} Paris Diderot; 5 place Jules Janssen, F-92195 Meudon, France\label{LUTh}
}

\date{Received ; accepted }

 
\abstract
{We present the discovery of two transiting extrasolar planets by the
  satellite \corot.} 
{We aim at a characterization of the planetary bulk parameters, which
  allow us to further investigate the formation and evolution of the
  planetary systems and the main properties of the host stars.}
{We used the transit light curve to characterize the planetary
  parameters relative to the stellar parameters.
  The analysis of HARPS spectra established the planetary nature of the
  detections, providing their masses.    
  Further photometric and spectroscopic ground-based observations
  provided stellar parameters (\logg,\teff,\vsini) to characterize the 
  host stars. 
  Our model takes the geometry of the transit to constrain the stellar
  density into account, which when linked to stellar evolutionary
  models, determines the bulk parameters of the star.   
  Because of the asymmetric shape of the light curve of one of the
  planets, we had to include the possibility in our model that the
  stellar surface was not strictly spherical.} 
{We present the planetary parameters of 
  CoRoT-28b, a Jupiter-sized planet 
  (mass $0.484\pm0.087\Mjup$; radius $0.955\pm0.066\Rjup$) 
  orbiting an evolved star with an orbital period of 
  $5.208\,66\pm0.000\,34$ days, 
  and CoRoT-29b, another Jupiter-sized planet 
  (mass $0.85\pm0.20\Mjup$; radius $0.90\pm0.16\Rjup$)
  orbiting an oblate star with an orbital period of 
   $2.850\,548\,7\pm0.000\,004\,1$ days. 
  The reason behind the asymmetry of the transit shape is not
  understood at this point.} 
{These two new planetary systems have very interesting properties and
  deserve further study, particularly in the case of the star CoRoT-29.}


\maketitle

%
\section{Introduction}
\label{section:introduction}

Spaceborne surveys of transiting extrasolar planets like
\corot\,\citep{baglin2006} or Kepler \citep{borucki2010a} have
provided crucial evidence of the interactions between stars and
planets and their common evolution.
In this paper we report the discovery and characterization of two
systems with interesting properties.

CoRoT-28b is a hot Jupiter orbiting an evolved star. 
We investigated the stellar and planetary properties, in particular
the rotation and metallicity of the star, and found a lithium
abundance  higher than expected for its evolutionary state. 

CoRoT-29b is a bit of a riddle. 
The transit discovered by \corot\,is significantly asymmetric, a
signature that the host star could have a non-spherical shape.
The asymmetry measured by \corot\,has been confirmed by
ground-based observations at several observatories.
We developed a code that successfully models the transit light
curve assuming that the star is oblated.
The deformation of the star implied by our model, however, is far too
large compared to our current understanding of stellar interiors.
Other possible scenarios are bands of stellar spots on the
surface of the star, which if properly placed, could also reproduce
the transit light curve measured.
We show the observational evidence and our modelling efforts, 
but we leave open the question of the origin of the asymmetry.

%
\section{CoRoT observations and data reduction}
\label{section:corot_observations}

\subsection{General description}
\label{subsection:corot_general_description}

The satellite \corot\,observed the field LRc08 (pointing coordinates
18h 28m 34.7s, 5\degr 36\arcmin 0.0\arcsec) in 2011 between 6 July
and 30 September. 
The \corot\,data for the run LRc08 has been public since 7 June 2013
and can be obtained through the IAS CoRoT 
Archive\footnote{{\texttt http://idoc-corot.ias.u-psud.fr/}}.

\begin{figure*}
  \centering
  \begin{minipage}[t]{0.48\textwidth}
    \includegraphics[%
    width=\linewidth,%
    height=0.5\textheight,%
    keepaspectratio]{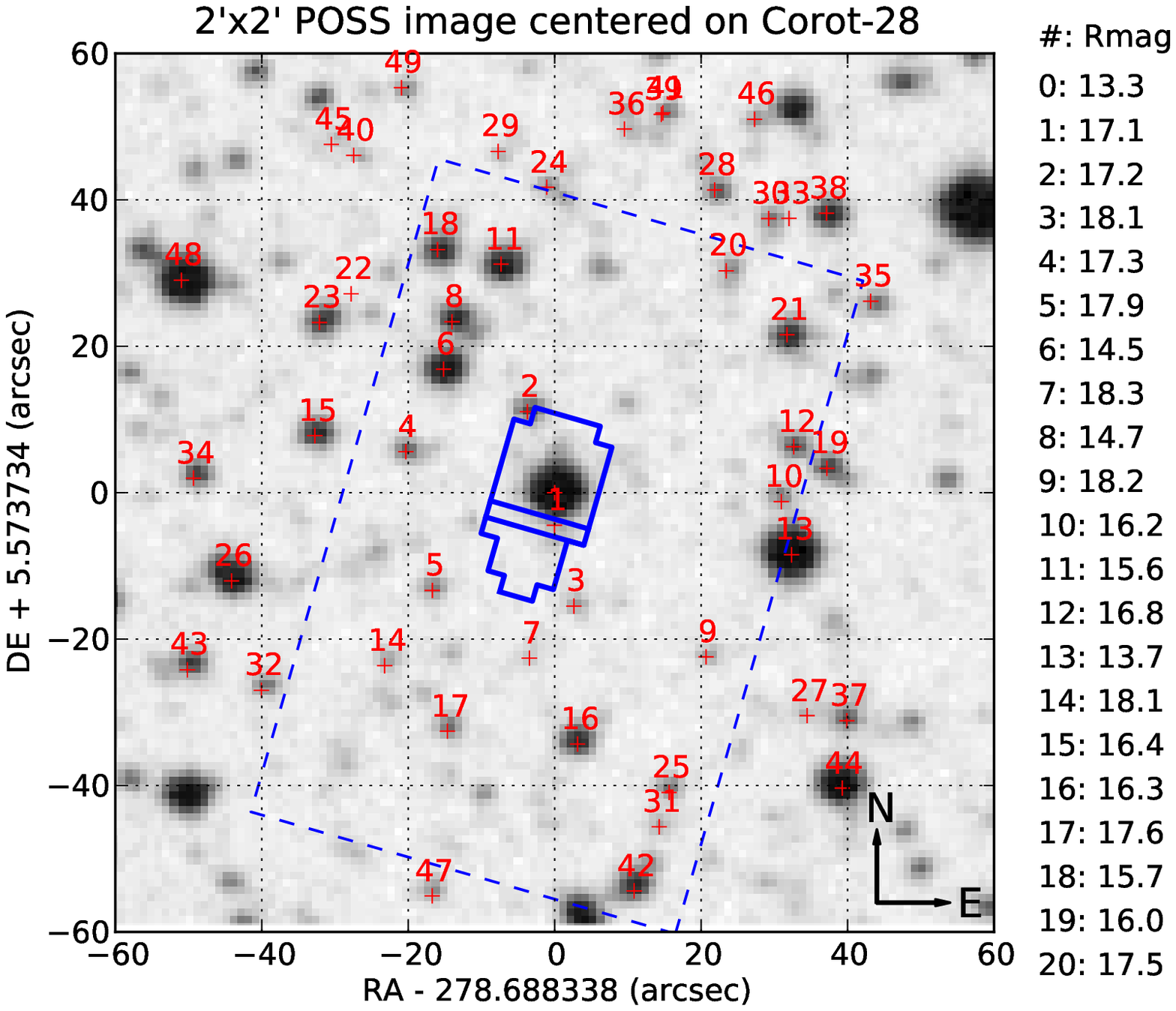}
  \end{minipage}
  \begin{minipage}[t]{0.48\textwidth}
    \includegraphics[%
    width=\linewidth,%
    height=0.5\textheight,%
    keepaspectratio]{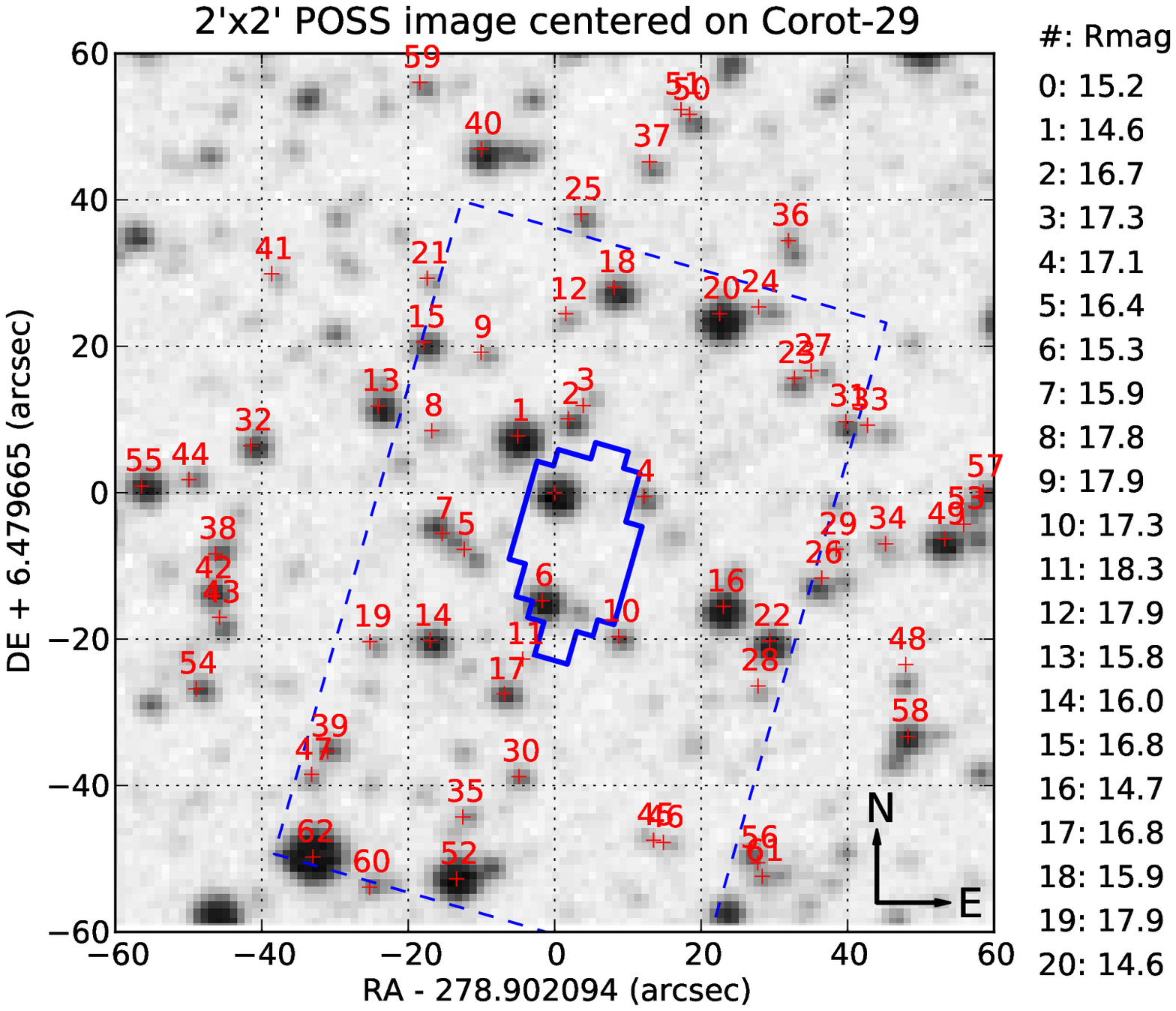}
  \end{minipage}
    \caption{Second Palomar Observatory Sky Survey (POSS II,
      \citealt{reid1991}) image around the CoRoT-28 (left) and
      CoRoT-29 (right) targets. 
      We have over-plotted the \corot\,masks used in the observations
      and we give identification labels for the brightest, closest 
      background stars.
      The mask of CoRoT-28 shows the subdivision in the three submasks,
      named from north to south: red, green, and blue.}
    \label{figure:masks}
\end{figure*}

The coordinates, identification labels, and magnitudes of the stars
CoRoT-28 and CoRoT-29 are given in Table~\ref{table:coordinates}.
CoRoT-28 is a relatively bright target and it was assigned an
observing cadence rate of 32s from the start of the observations,
collecting $225\,087$ measurements with a chromatic mask. 
CoRoT-29 is a fainter target and was assigned the cadence rate of 512s. 
When the first transits were discovered by the Alarm Mode, the
target was assigned a sampling rate of 32s, collecting in total
$121\,288$ measurements with a monochromatic mask.
The region around both targets, indicating the position of the nearest
contaminants and the orientation of the masks, is shown in
Fig.~\ref{figure:masks}.
For an overview of the \corot\,observing modes, please refer to
\citet{boisnard2006,barge2008b,auvergne2009}.

\begin{table}
\caption{IDs, coordinates, and magnitudes of CoRoT-28 and CoRoT-29.}
\label{table:coordinates}      
\centering
\begin{tabular}{lcc}       
\hline\hline                 
CoRoT name      & CoRoT-28b           & CoRoT-29b           \\
window ID       & LRc08~E2~0275       & LRc08~E2~3502       \\
CoRoT ID        & 652180991           & 630831435           \\
USNO-A2 ID      & 0900-13187173       & 0900-13220969       \\ 
USNO-B1 ID      & 0955-0374996        & 0964-0373597        \\ 
2MASS ID        & 18344520+0534254    & 18353650+0628467    \\
\\
\multicolumn{3}{l}{Coordinates} \\
\hline         
RA (J2000)  & 18h 34m 45.21s       & 18h 35m 36.50s        \\
Dec (J2000) & 5\degr 34\arcmin 25.72\arcsec & 6\degr 28\arcmin 46.99\arcsec \\
\\
\multicolumn{3}{l}{Magnitudes} \\
\hline
\centering
Filter & \multicolumn{2}{c}{values} \\
\hline
B\tablefootmark{a}  & 14.994 $\pm$ 0.033   & 16.704 $\pm$ 0.105   \\
V\tablefootmark{a}  & 13.775 $\pm$ 0.022   & 15.560 $\pm$ 0.069   \\
r'\tablefootmark{a} & 13.246 $\pm$ 0.012   & 15.090 $\pm$ 0.060   \\
i'\tablefootmark{a} & 12.627 $\pm$ 0.036   & 14.515 $\pm$ 0.025   \\
J\tablefootmark{b}  & 11.601 $\pm$ 0.023   & 13.565 $\pm$ 0.021   \\
H\tablefootmark{b}  & 11.143 $\pm$ 0.025   & 13.120 $\pm$ 0.021   \\
K\tablefootmark{b}  & 11.032 $\pm$ 0.025   & 13.048 $\pm$ 0.031   \\
W1\tablefootmark{c} & 10.912 $\pm$ 0.022   & 12.974 $\pm$ 0.030   \\
W2\tablefootmark{c} & 10.987 $\pm$ 0.022   & 13.064 $\pm$ 0.036   \\
W3\tablefootmark{c} & 11.866 $\pm$ 0.314   & 12.088 $\pm$ 0.295   \\
W4\tablefootmark{c} &  9.031 (upper limit) &  9.031 (upper limit) \\
\hline
\end{tabular}
\tablefoot{
  \tablefoottext{a}{Provided by ExoDat \citep{deleuil2009};}
  \tablefoottext{b}{from 2MASS catalogue;}
  \tablefoottext{c}{WISE photometry from \citet{wright2010wise}.}
}
\end{table}

\subsection{CoRoT-28b}
\label{subsection:corot_reduction_corot28}

\begin{figure}
  \centering
  \includegraphics[%
  width=0.9\linewidth,%
  height=0.5\textheight,%
  keepaspectratio]{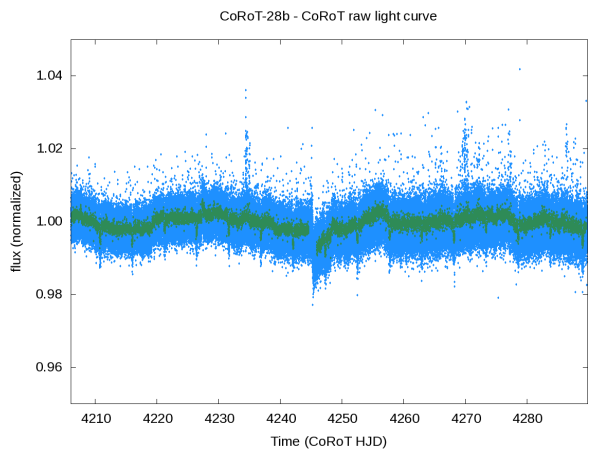}
  \caption{Raw light curve of CoRoT-28. 
    The original data points, sampled at 32s, are shown in blue. 
    We have binned the data to 512s bins to guide the eye.} 
  \label{figure:corot28_raw}
\end{figure}

The raw light curve of CoRoT-28 is shown in
Fig.~\ref{figure:corot28_raw}.
We treated the EN2 level\footnote{For more information
  about the \corot\,data products, please see \citet{baudin2006}.}
light curve of CoRoT-28 observed by \corot\,in its version 3.0. 
The removal of the influence of instrumental systematics and stellar
variability is a necessary step before obtaining the planetary
parameters from the light-curve fit.
Standard methods consist in the subtraction of a low-level polynomial
(in our case, a parabola) to an interval around each transit with a
length of a few transit durations (in our case, three). 
This method was applied successfully with the light curve of
CoRoT-28, and we then proceeded to the modelling of the planetary
parameters as described in
subsection~\ref{subsection:planetary_parameters_c28b}.

\subsection{CoRoT-29b}
\label{subsection:corot_reduction_corot29}

In the case of CoRoT-29, the standard reduction procedures described
in Section~\ref{subsection:corot_reduction_corot28} revealed a feature 
with the shape of non-zero slope in the bottom of the transit shape,
which is otherwise expected to be reasonably flat.
The slope of the bottom part of the transit is non-zero with 95\%
confidence level in the 32s sampled data by \corot, and only non-zero
with 1$\sigma$ significance in the 512s sampled data set. 
But at the same time, CoRoT-29 is a fainter star than CoRoT-28 and its
light curve has more noise at the timescales comparable with the 
transit length. 
We therefore decided to use a more robust filtering for this
light curve.

We followed the procedure described in \citet{alapini2009}, which
iteratively fits the planetary parameters, subtracts the planetary
model from the data, creates a model of the stellar variability, then
subtracts the stellar variability model from the original data, and
then fits the planetary parameters again until the method converges
to a final set of planetary parameters and stellar activity pattern.  
In the first iteration, we used a sliding time window of 0.5 days in
length to deal with the variations due to activity. 
For the interpolation, we removed the points measured during the
transit. 
But for the next iteration steps we divided the light curve in
intervals of 0.5 days and fitted each interval with a Legendre
polynomials of order 5.
This was chosen because a sliding window with short timescales might
have an influence on sharp features in the light curve, such as the
transit ingress or egress, and the method might fail to converge
\citep[see also the discussion in][]{bonomo2012b}.

\begin{figure}[th!]
  \centering
  \includegraphics[%
  width=\linewidth,%
  height=0.5\textheight,%
  keepaspectratio]{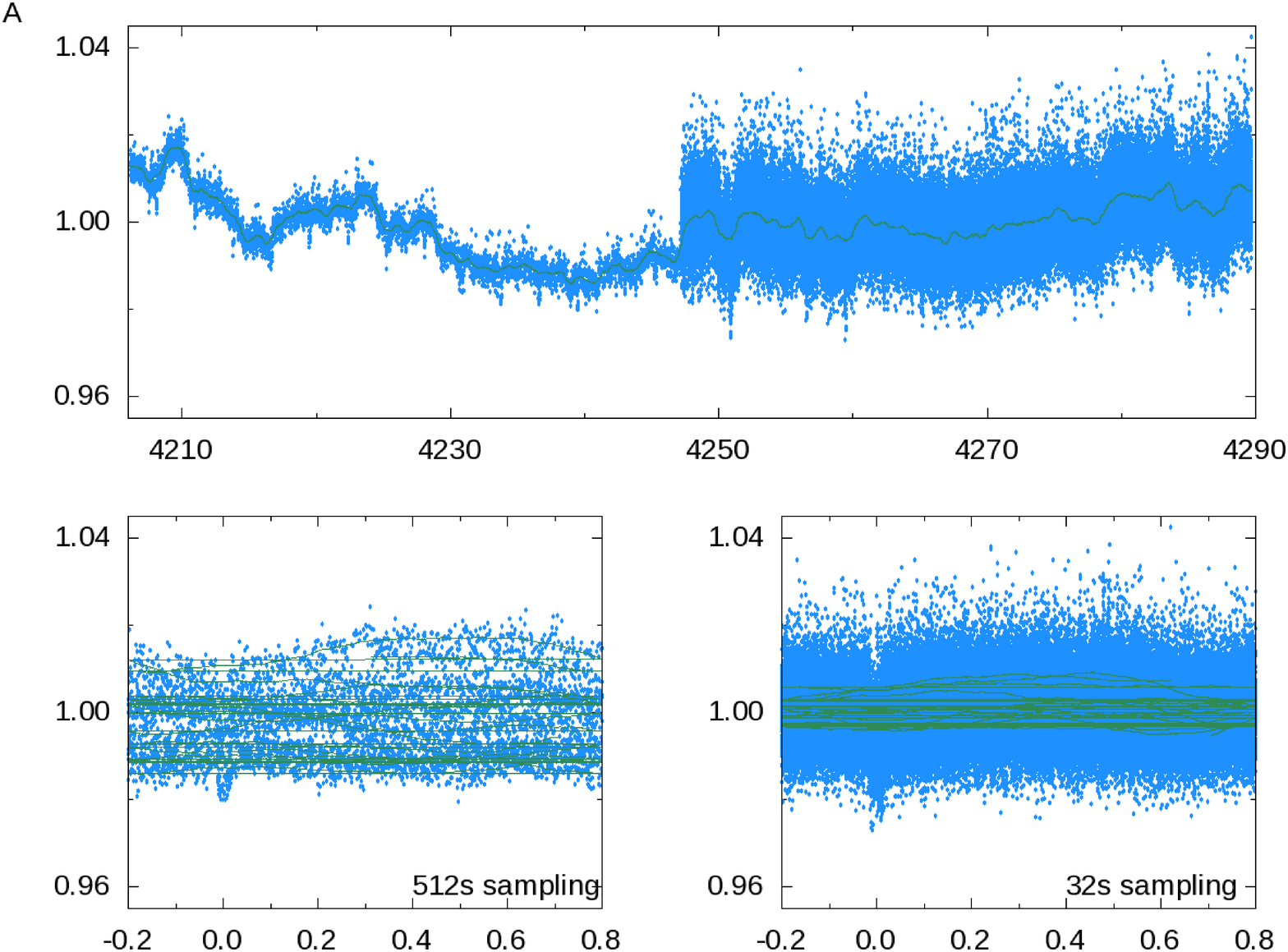}
  \includegraphics[%
  width=\linewidth,%
  height=0.5\textheight,%
  keepaspectratio]{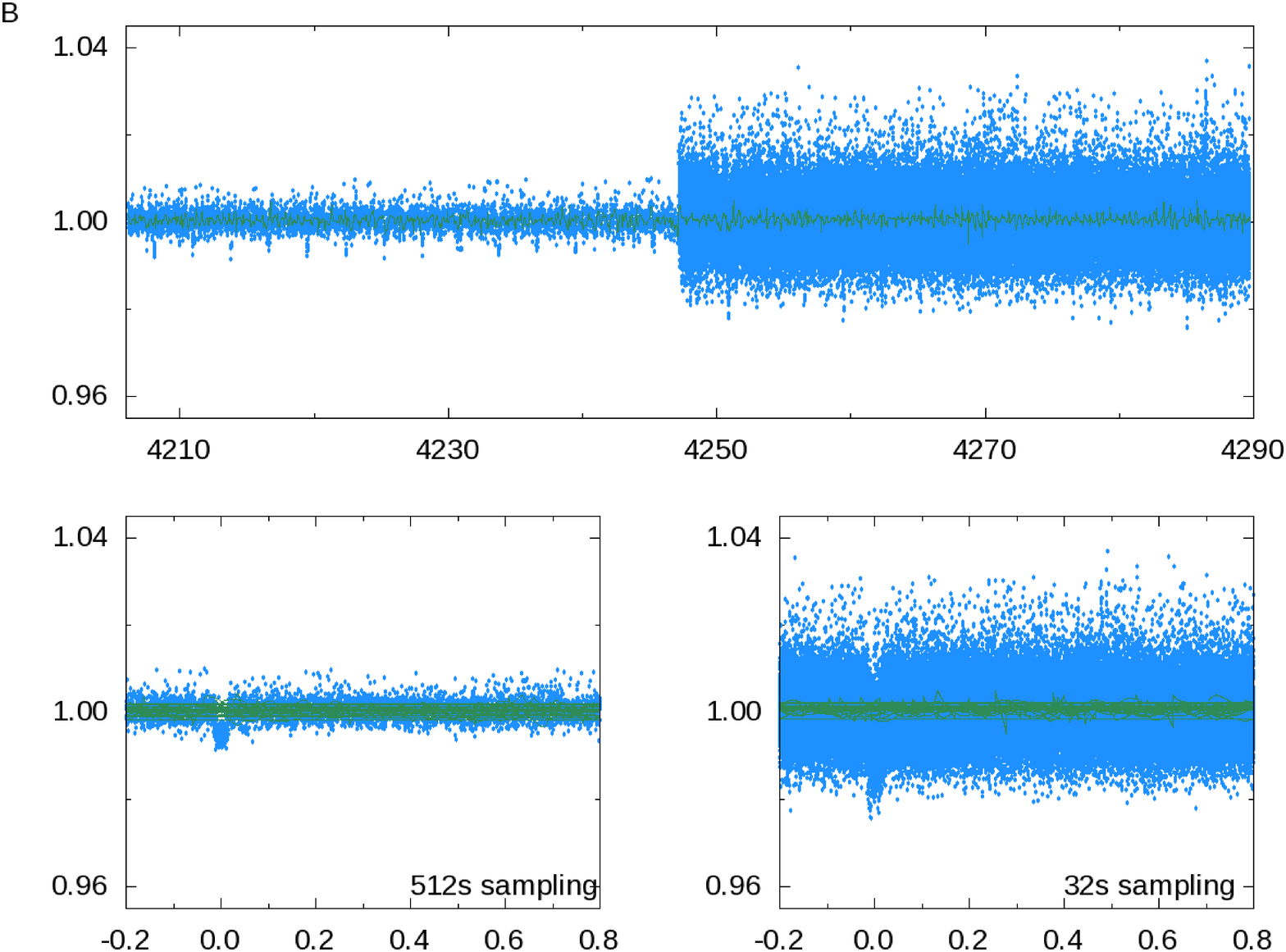}
  \includegraphics[%
  width=\linewidth,%
  height=0.5\textheight,%
  keepaspectratio]{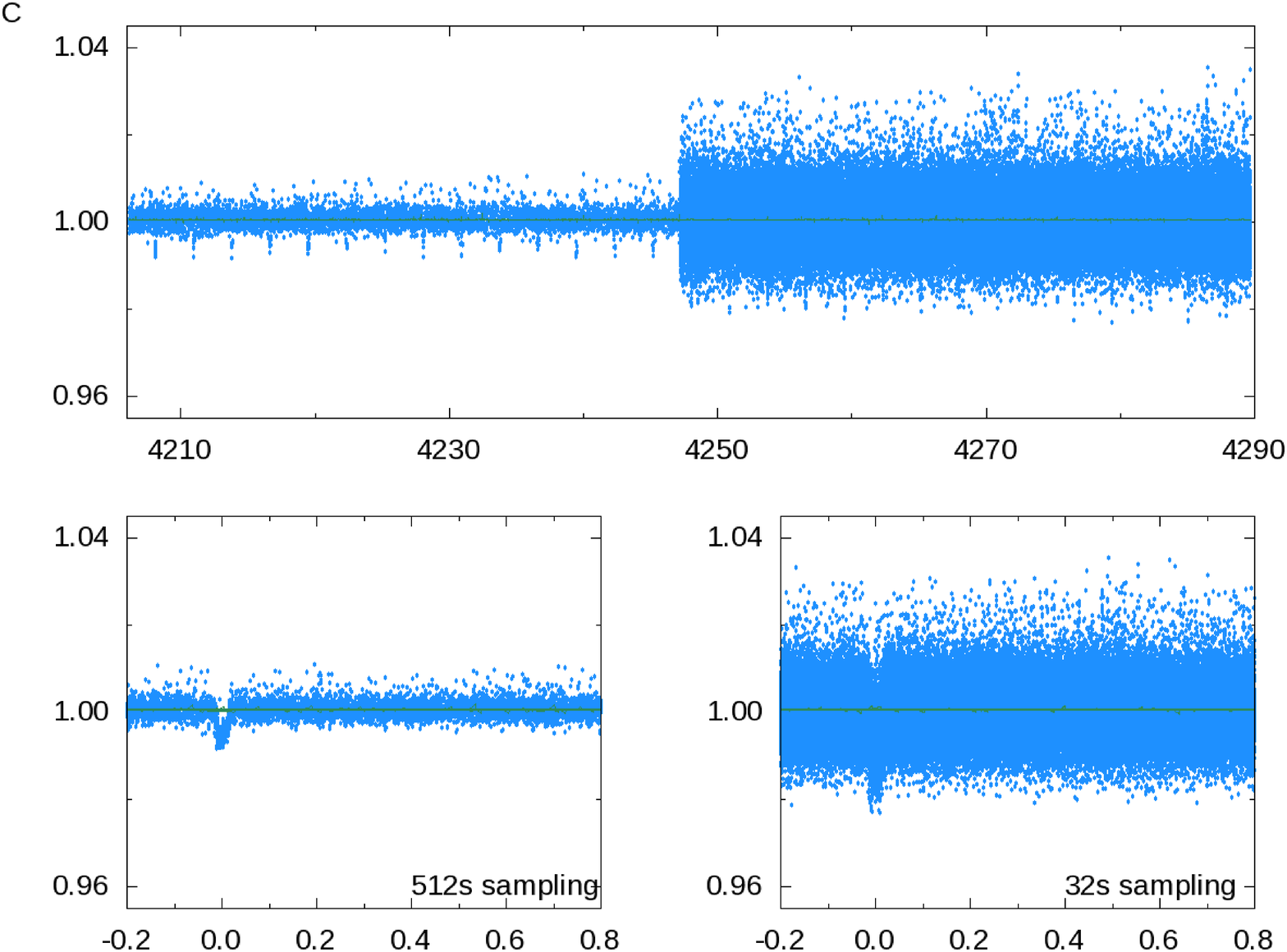}
  \caption{Iterative filtering applied to the light curve of
    CoRoT-29b, converging in three steps (labelled A, B, and C). 
    Top of each panel shows the light curve (normalized flux vs CoRoT
    Julian Date, which is HJD - 2451545.0), the bottom figures on each
    panel show the folded light curves (normalized flux vs orbital
    phase).} 
  \label{figure:CoRoT29bFiltering}
\end{figure}

We tried with intervals from 0.5 to 1.5 days in steps of 0.1 days and
though the removal of the systematics is not equally satisfactory for
each interval, the non-zero slope in the bottom transit of CoRoT-29b
remained.
A Legendre polynomial of order 5 within a 0.5 day window provides a 
smooth fit to the data and it is able to remove the contributions from 
the out-of-transit, spot-induced stellar activity and low-frequency
instrumental systematics present in the data. 
The regions of the light curve sampled at 512s and at 32s were
treated as independent data sets.
Our method converged after three iterations (one with the sliding window
and two with the Legendre polynomials, see
Fig.~\ref{figure:CoRoT29bFiltering}).

\subsection{Background correction of CoRoT-29b}
\label{subsection:corot_background_corot29}

\begin{figure}
  \centering
  \includegraphics[%
  width=0.9\linewidth,%
  height=0.5\textheight,%
  keepaspectratio]{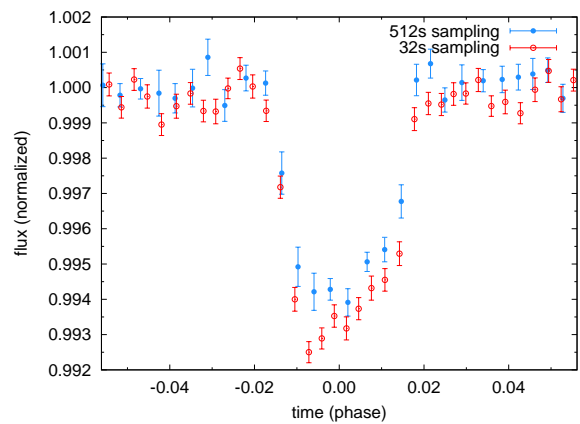}
  \caption{Change in the transit depth of CoRoT-29b observed in the 
    light curves sampled at 512s and 32s (version 3.0 of the data).}
  \label{figure:CoRoT29b_change_tr_depth}
\end{figure}

Comparing the transits of the section of the \corot\,light curve
sampled at 512s and at 32s, we were able to identify a change in the
observed transit depth, which required further study (see
Fig.~\ref{figure:CoRoT29b_change_tr_depth}). 
\citet{auvergne2009} described the photometric capabilities of the
\corot\,satellite in detail.  
However some of those capabilities have slightly degraded with
the ageing of the instrument.
One effect is the increase of the dark current of the CCD, which has
an important impact on the on-board photometry.  
The dark current of the CCD was modelled as a uniform value calculated
in 196 (14x14) windows distributed along the CCD: 3/4 of them sampled
at 512s and 1/4 of them sampled at 32s.
The correction applied to the 32s and 512s light curves was the median
value of the 32s and 512s background windows, respectively, to mitigate
the impact of hot pixels and cosmic ray impacts.
However, with the ageing of the CCD, the uniform model for the dark
current is no longer a faithful description of the phenomenon.
We observed a gradient along the Y axis of the CCD whose
slope has increased with time.
Moreover, the distribution of background windows sampled at 512s and
32s is not uniform, which introduces an additional bias in the
background correction.
The median background level at 32s sampling is calculated in a region 
of the CCD that has in average a lower Y value than the average of
the windows sampled at 512s, which introduces a visible difference 
in the mean level of the light curve at the two different sampling
rates.
This effect is common to all light curves on the CCD during this
particular run, but it cannot be seen in the light curve of
CoRoT-28, as this target was only sampled at the 32s cadence.
This effect can be mitigated by modifying the corresponding background
correction for each star as a function of its position on the CCD, a
feature that will be introduced in future releases of \corot\,data.

The \corot\,pipeline is currently implementing an improved background
correction to mitigate this effect.
We used the data of this improved pipeline (version
3.5\footnote{The version 3.0 of the data is already publicly accessible
  through the IAS CoRoT Archive 
  ({\texttt http://idoc-corot.ias.u-psud.fr/}). 
  The version 3.5 of the data will be publicly available in coming months.}).
However, the final planetary parameters determined when using the
previous (version 3.0) or the new data (version 3.5) are not
significantly different. 
Our modelling fits the background contamination as a free parameter,
and can deal with the residuals left by the background correction from
the pipeline. 
The incomplete background correction pipeline dilutes the transit
signal as if there was an additional contribution of light from a
background star, a well-known effect that can be corrected with
standard procedures.

%
\section{Ground-based observations}
\label{section:groundbased_observations}

The main objective of ground-based photometric follow-up is to check
whether the observed transit features occur on the target star or
might arise on a nearby eclipsing-binary system \citep{deeg2009}.
Then, ideally, low-resolution spectroscopic measurements characterize 
the nature of the stellar target before high-resolution radial
velocity measurements independently confirm the planetary nature of
the target providing a value of its mass \citep{moutou2009a}.
This section describes the ground-based observations carried out to
characterize these \corot\,targets. 

\subsection{Photometric measurements}
\label{subsection:photometric_measurements}

\noindent {\it CoRoT-28b}\\\mbox{ }

Photometric follow-up of CoRoT-28b was performed with the IAC80
(Tenerife) and the Euler (La Silla) telescopes. 
Photometric data were acquired at the 80cm telescope of the IAC on
Tenerife (IAC80). 
Measurements during the transit (on) obtained on 9 October 2011 and
out-of-transit (off) data taken on 25 October 2011 showed an on-off
brightness variation of 0.3\%, but at a low level of confidence.
Each observing run lasted 33 minutes.
Conversely, relevant brightness variations could be excluded with
high confidence on the neighboring stars (stars 2, 3, 6, 13, and 16
in Fig.~\ref{figure:masks}).
A star some 17\arcsec SSW of the target, on which photometry was
performed, was found to be 5.8 mag fainter than the target and hence
too faint to cause a false alarm; the recognizable stars that were
closest to the target were even fainter (about 4\arcsec N and 
5\arcsec S; no photometry could be performed on them). 
In the data from the Euler telescope, taken in 2011 on 29 August (on)
and on 2 September (off), no eclipses were found, neither
on-target nor on any nearby stars. 
Given that the uncertainty in the ephemeris at the time of both
observations was only 4 minutes, we can therefore conclude that the
transits must occur on the target.

\mbox{ }\\\noindent {\it CoRoT-29b}\\\mbox{ }

\begin{figure}
  \centering
  \includegraphics[%
  width=0.9\linewidth,%
  height=0.5\textheight,%
  keepaspectratio]{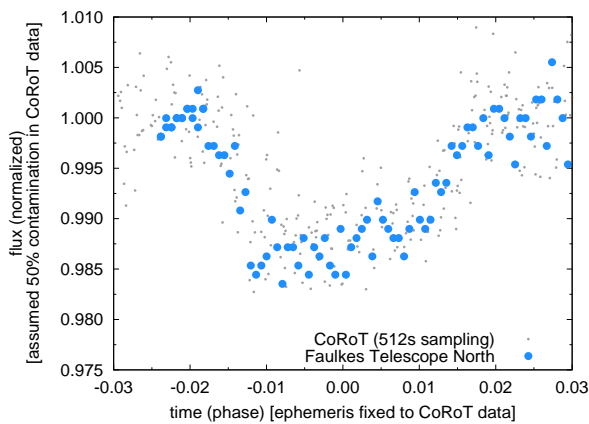}
  \caption{Superposition of the transits of CoRoT-29b observed by
    \corot\,(small gray points) in 2011 and the transit observed by
    FTN in 2012 (large blue points). 
    This is not a fit. 
    In this Figure we just superpose the FTN and the \corot\,data
    corrected from the contamination measured in the light curve.
    The contamination value was obtained exclusively from the fit to
    the \corot\,data.
    See text for discussion.}
  \label{figure:ftn}
\end{figure}

Photometric follow-up observations of CoRoT-29b were performed on
several occasions. 
A full transit was obtained on 28 May 2012 with the 2m Faulkes
Telescope North (FTN).
The analysis of this data showed a clear transit on CoRoT-29 itself.
There are several nearby stars which fall within the same PSF and
aperture in the \corot\,data, contaminating flux of the mask of
CoRoT-29.
Most of this contamination is due to a similarly bright star at a
distance of 10\arcsec NW which can be well separated in the ground
based images. 
Therefore the real depth of the transit, as it is observed from ground
is 1.5\% (measured from FTN observations) instead of the 0.6\% in the
\corot\,light curve, because the contaminating star accounts for about
50\% of the flux in the \corot\,mask (see
Table~\ref{table:stellar_planetary_parameters}). 
The FTN observations show also the non-flat feature at the bottom of
the transit observed by \corot\,(see Fig.~\ref{figure:ftn}).
Given the on-target detection of the transit, contamination from
eclipsing binaries at distances larger than 2\arcsec could be excluded
as a source of false alarm. 

Further ON-OFF photometry undertaken by the 3.5m CFHT on 10 May 2013
also showed an on-target transit with a depth of $1.4 \pm 0.3\%$.  

\begin{figure}
  \centering
  \includegraphics[%
  width=0.9\linewidth,%
  height=0.5\textheight,%
  keepaspectratio]{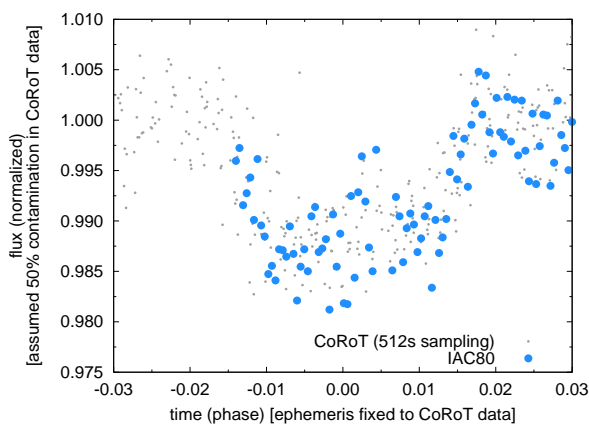}
  \caption{Superposition of the transits of CoRoT-29b observed by
    \corot\,(small gray points) in 2011 and the transit observed by
    IAC80 in 2014 (large blue points). Same conditions as in
    Fig.~\ref{figure:ftn}.
    }
  \label{figure:iac80}
\end{figure}

A full transit of CoRoT-29b was observed with IAC80 in the night
of 14 July 2014. 
The transit centre time, at BJD $2\,456,853.435 \pm 0.005$, was used
to refine the ephemeris in
Table~\ref{table:stellar_planetary_parameters}. 
It also shows some degree of asymmetry, though it is difficult to
assess it with the same confidence level as for FTN data 
(see Fig.~\ref{figure:iac80}).
The ingress is partially missing, complicating the assessment of any
possible out-of-transit slope, either caused by stellar activity or
by residuals of the data reduction (airmass correction,
instrumentals, etc.).

\subsection{Radial velocity measurements}
\label{subsection:radialvelocity_measurements}

\noindent {\it CoRoT-28b}\\\mbox{ }

Radial velocity measurements were first obtained with the SOPHIE 
\citep{perruchot2008,bouchy2009a} spectrograph on the 193 cm
telescope at the Observatoire de Haute Provence (France) and showed
variations in phase with the \corot\,ephemeris, establishing the 
planetary nature of the candidate. 
A total of 25 radial velocity measurements were acquired in High
Efficiency mode ($R=40\,000$) with SOPHIE between 27 August 2011 and 
17 August 2012. 
An additional seven measurements between 7 July 2012 and 10 August 2013 were
performed using the HARPS spectrograph in standard HAM mode 
($R=115000$) \citep{mayor2003} mounted on the 3.6 m ESO telescope at
La Silla Observatory (Chile) as part of the ESO large program
188.C-0779.
Finally, ten spectra of CoRoT-28 were taken at different epochs between
June and July 2012 using the FIbre-fed {\'E}chelle Spectrograph
\citep[FIES;][]{frandsen1999,telting2014} mounted at the 2.56-m Nordic
Optical Telescope (NOT) of Roque de los Muchachos Observatory (La
 Palma, Spain).
We employed the \emph{high-res} fibre, which provides a resolving power
of R\,=\,67\,000 in the spectral range 3600\,--\,7400\,\AA.

The SOPHIE and HARPS spectra were extracted using the respective
pipelines.
The radial velocities were then computed following a technique of
weighted cross-correlation of the spectra using a numerical G2 star
mask. 
This technique is described by \cite{baranne1996} and
\cite{pepe2002}. 
The spectral orders with low signal-to-noise ratio were discarded to
perform the computation to reduce the dispersion of the measurements. 
We eliminated the two bluest and five reddest orders for SOPHIE (39 in
total), the ten bluest and the two reddest for HARPS (72 in total). 
The cross-correlation function of CoRoT-28b shows a single peak with
FWHM of 9.9 \kms\, with SOPHIE and 7.6 \kms\, with HARPS. 
Also, SOPHIE and HARPS radial velocities obtained with different
numerical spectral masks (F0, G2 and K5) show the same behaviour,
suggesting that the radial velocity variations cannot be explained by
a blend scenario with different spectral type stars. 

For the FIES data, we adopted the same observing strategy described in 
\citet{buchhave2010}, i.e. we split each epoch observation into three
consecutive exposures of 1200~sec to remove cosmic rays hits, and
we acquired long-exposed ThAr spectra in ``sandwich-mode'' to trace
the RV drift of the instrument. 
We reduced the data  using a customized IDL software suite, which
includes bias subtraction, flat fielding, order tracing and
extraction, and wavelength calibration. 
We obtained radial velocity measurements  performing a multi-order
cross-correlation with the RV standard star HD\,182572, observed with
the same instrument set-up as the target object, and for which we
adopted an heliocentric radial velocity of $-100.350$\,\kms, as
measured by \citet{udry1999}.   

\begin{figure}
  \centering
  \includegraphics[%
  width=0.9\linewidth,%
  height=0.5\textheight,%
  keepaspectratio]{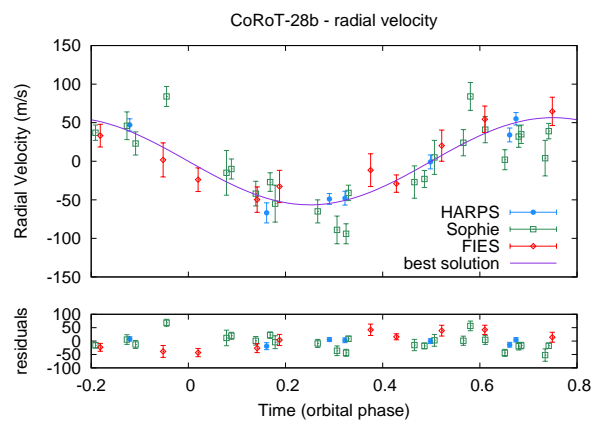}
  \caption{Radial velocity measurements and best solution for
    CoRoT-28. Offsets have been subtracted according to the values
    given in Table~\ref{table:stellar_planetary_parameters}.}
  \label{figure:corot28_rv}
\end{figure}

\begin{figure}
  \centering
  \includegraphics[%
  width=0.9\linewidth,%
  height=0.5\textheight,%
  keepaspectratio]{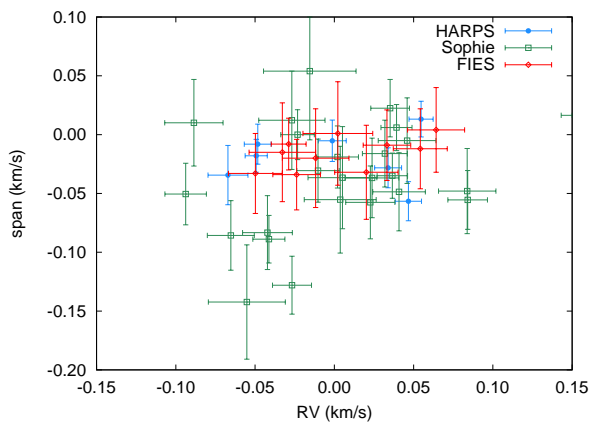}
  \caption{Bisector span as a function of radial velocities for
    CoRoT-28. 
    We  subtracted the systemic velocities in the X axis. 
    The uncertainties in the bisector span is assumed to be twice the 
    uncertainties in the radial velocities.
    The bisector span shows no correlation with the radial velocities.}
  \label{figure:corot28_bis}
\end{figure}

The radial velocity measurements with the best orbital solution are
shown in Figure~\ref{figure:corot28_rv}, assuming the period and
transit epoch determined by the CoRoT light curve.  
Finally, the cross-correlation function bisector span shows no
correlation with the radial velocities 
(see Fig.~\ref{figure:corot28_bis}), 
reinforcing the conclusion that the radial velocity variations are not
due to a blending effect.
Radial velocities, their errors, and bisector span are listed in
Tables~\ref{table:rv_corot-28_SOPHIE} to 
\ref{table:rv_corot-28_FIES}.  

For CoRoT-28b, we fit SOPHIE, HARPS, and FIES data  together while
fixing the ephemeris derived from the light-curve analysis (see
Table~\ref{table:stellar_planetary_parameters}).
The rest of the orbital parameters (orbital
eccentricity, argument of periastron, radial velocity semi-amplitude,
and systemic velocities) were derived using the modelling procedure and
Markov Chain Monte Carlo module from PASTIS \citep{diaz2014}. 
The priors used for the orbital parameters were non-informative
(uniform priors) except for the period and the transit epoch that were
fixed. 
The results are summarized in
Table~\ref{table:stellar_planetary_parameters}. 

\mbox{ }\\\noindent {\it CoRoT-29b}\\\mbox{ }

\begin{figure}
  \centering
  \includegraphics[%
  width=0.9\linewidth,%
  height=0.5\textheight,%
  keepaspectratio]{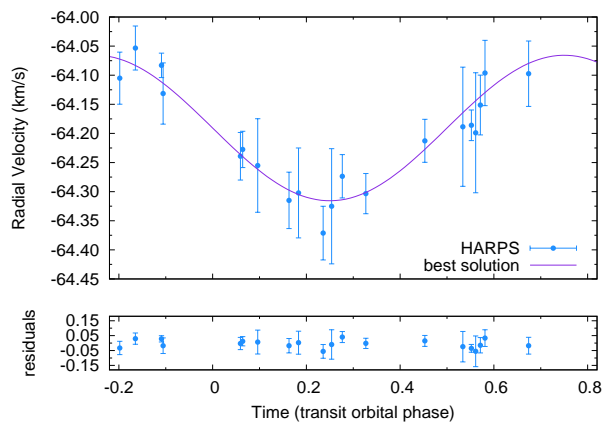}
  \caption{Radial velocity measurements and best solution for
    CoRoT-29. Offsets were subtracted according to the values
    given in Table~\ref{table:stellar_planetary_parameters}.}
  \label{figure:corot29_rv}
\end{figure}

\begin{figure}
  \centering
  \includegraphics[%
  width=0.9\linewidth,%
  height=0.5\textheight,%
  keepaspectratio]{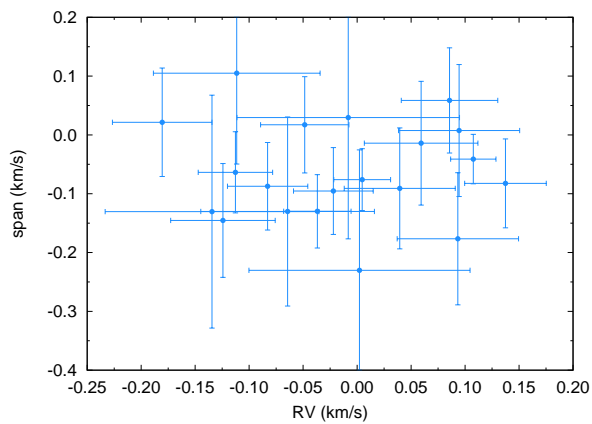}
  \caption{Bisector span as a function of radial velocities for
    CoRoT-29.     
    We  subtracted the systemic velocities in the X axis. 
    The uncertainties in the bisector span is assumed to be twice the 
    uncertainties in the radial velocities.
    The bisector span shows no correlation with the radial velocities.}
  \label{figure:corot29_bis}
\end{figure}

Because of its faintness ($V=15.6$), CoRoT-29 radial velocity follow-up
could only be performed with HARPS in the EGGS mode to improve the
throughput. 
Compared to the HAM mode, which was used to follow up CoRoT-28, the EGGS
mode of HARPS has a larger fiber ($1.4 \arcsec$ compared to $1
\arcsec$) and no scrambler. 
The spectral resolution is therefore reduced ($R=80\,000$) but the
throughput is twice as large. 
We obtained 20 measurements  between 19 June 2012 and 10 August 2013,
showing that the radial velocity variations were in agreement 
with the CoRoT ephemeris and establishing the planetary nature of 
CoRoT-29b (see Fig.~\ref{figure:corot29_rv}). 
The same cross-correlation technique with a G2 spectral type numerical
mask allowed us to derive the radial velocities. 
As for CoRoT-28, the ten bluest and the two reddest orders were
discarded. 
We obtained a 7.1 \kms\, single peak cross-correlation function. 
Also, cross-correlation with other masks (F0 and K5) show no
discrepancies in the radial velocity variations. 
Moreover, the cross-correlation function bisector span shows no
correlation with the radial velocities (see
Figure~\ref{figure:corot29_bis}). 
This suggests that the probability to have a blending effect is
very low.  
The radial velocity measurements are represented on
Figure~\ref{figure:corot29_rv} and are listed with their errors and
bisector spans in Table~\ref{table:rv_corot29_HARPS} .

We also fit CoRoT-29b HARPS data while fixing the ephemeris as
described above for CoRoT-28b and the results are also summarized in 
Table~\ref{table:stellar_planetary_parameters}. 

\subsection{Spectroscopic characterization}
\label{subsection:spectroscopic_measurements}

To derive the stellar parameters from the spectroscopic data,
we used the Spectroscopy Made Easy (SME) package (version 5, February
2014).  
The development and structure and performance of SME is described in
\citet{valenti1996} and in e.g. \citet{valenti2005}. 
Briefly, this code uses a grid of stellar models to which is fitted
the observed spectrum. 
This is done by calculating a synthetized spectrum and minimizing the
discrepancies through a non-linear, least-squares algorithm. 
The analysis is based on the latest generation of MARCS model
atmospheres \citep{gustafsson2008} and the ATLAS12 models
\citep{kurucz2005}. 
The models calculate the temperature and pressure distribution in
radiative and hydrostatic equilibrium, assuming a plane-parallel
stellar atmosphere. 

The SME is developed so that, by matching synthetic spectra to observed
line profiles, one can extract the information in the observed
spectrum and search among stellar and atomic parameters until the best
fit is achieved. 
The SME utilizes automatic parameter optimization using a
Levenberg-Marquardt \chisq\, minimization algorithm. 
Synthetic spectra are calculated on the fly, by a built-in spectrum
synthesis code, for a set of global model parameters and specified
spectral line data. 
Starting from user-provided initial values, synthetic spectra are
computed for small offsets in different directions for a subset of
parameters defined to be “free”. 
The model atmospheres required for this are calculated through
interpolating the grid of models mentioned above. 

We used as input data the high-resolution spectra taken with HARPS
(see Section~\ref{subsection:radialvelocity_measurements}).
In the observed spectra ,we use a large number of spectral lines,
e.g. the Balmer lines (the extended wings are used to constrain
\teff), \ion{Na}{I} D lines, \ion{Mg}{I} b, and \ion{Ca}{I}
(for \teff\,and \logg) and a large number of metal lines
(to constrain the abundances). 
For CoRoT-28, we obtained a \ion{Li}{I} equivalent width of
31m\angstrom\,at $6\,708$\angstrom.

After finding the \teff\,from the Balmer line wings 
\citep[e.g.][]{fuhrmann1998}, we use the calibration of
\citet{bruntt2010a} to fix the micro- and macro-turbulence parameters
before finding other parameters.
The quoted errors are 1 $\sigma$ standard deviation errors in SME and
thus a measure of how well the generation of the synthetic spectrum
has succeeded. 
We know (Fridlund et al., in preparation) that when we apply SME to
our (only) known source, the Sun, we find somewhat larger errors such
as a $\pm 100$K difference in \teff, and a $\pm 0.1$dex in [Fe/H] (as
proxy for metallicity) depending on the choice of model grid being used
(e.g. Atlas12 or MARCS 2012). The mass, radius, and age of the star are calculated using stellar
evolution tracks from \citet{hurley2000} and the constraints from the
analysis of the light curve (see
Section~\ref{section:planetary_parameters}). 
They are reported in Table~\ref{table:stellar_planetary_parameters}.

Additionally, we estimated the interstellar extinction \Av\, and
spectroscopic distance $d$ to CoRoT-28 and CoRoT-29 following the
method described in \citet{gandolfi2008}.
We simultaneously fitted the available ExoDat, 2MASS, and WISE colours
(Table~\ref{table:coordinates}) with synthetic theoretical magnitudes
derived by \emph{NextGen} low-resolution model spectra
\citep{hauschildt1999a} having the same photospheric parameters as the
stars. 
We excluded the $W_3$ and $W_4$ WISE magnitudes because of the low S/N
ratio. 
Assuming a normal extinction ($\Rv=3.1$), we found that CoRoT-28 and
CoRoT-29 are at $560 \pm 70$\,pc and $765 \pm 50$\,pc from the Sun,
and suffer an interstellar extinction of $\Av=0.75 \pm 0.20$\,mag and
$\Av = 0.85 \pm 0.15$\,mag, respectively.
The dereddened B-V excess of the stars is $0.98 \pm 0.06$ and 
$0.87 \pm 0.03$ mag, respectively.

\subsection{Spectral typing}
\label{subsection:spectral_typing}

A spectral type was derived for CoRoT-28 from a spectrum taken on 24
July 2012, with the low-resolution spectrograph
($\frac{\lambda}{\Delta\lambda}\approx1000$) at the Nasmyth focus of
the 2m telescope in Tautenburg. 
A slit width of 1\arcsec and the V200 grism were used covering the
wavelength range 360-935~nm. 
The signal-to-noise of the spectrum is about 180. 
The spectrum was reduced and extracted with IRAF \citep{tody1986,tody1993}
in the standard way, including subtraction of the bias and the
background as well as flat-field and extinction correction. 
Wavelengths were calibrated using sky lines in the long-slit spectra
and exposures of He and Kr gas discharge lamps. 
The spectral type of CoRoT-28 was obtained following
\citet{gandolfi2008} and \citet{sebastian2012}.
Briefly, we fitted the observed low-resolution spectrum using template
spectra from the library by \citet{valdes2004} in combination with the
photospheric parameters derived by \citet{wu2011}. 
Although a best match was attained for G8/G9V templates, we also found
matching templates with spectral classes G5-K0 and luminosity classes
IV and III. 
The luminosity class is not well defined by the low-resolution
spectrum. 
For the reasons explained above and considering the transit
parameters, we consider that the star is slightly evolved and classify
it as G8/9IV.
For CoRoT-29, considering the effective temperature, we choose a
spectral type K0V following \citet{kenyon1995}.

%
\section{Planetary parameters}
\label{section:planetary_parameters}

\subsection{CoRoT-28b}
\label{subsection:planetary_parameters_c28b}

For the transit analysis and final planetary parameter determination
of CoRoT-28b, we used the light curve that was filtered for stellar
variability, as explained in
Section~\ref{subsection:corot_reduction_corot28}, and the
information from the spectral analysis described in 
Section~\ref{section:groundbased_observations}. 
Since no significant transit timing variations (TTVs) were found, we
folded the light curve using the ephemeris in 
Table~\ref{table:stellar_planetary_parameters}. 
The folded light curve was then modelled with the 
{\it Transit Light Curve Model} code, described in
\citet{csizmadia2011}. 
Each chromatic light curve (red, green, blue, and white, see
Section~\ref{subsection:corot_general_description}) was treated in
the same way. 
For the transit light-curve fit, we used the \citet{mandel2002} model. 
The optimization process consists of two steps. First, the best
solution is found with a Harmony Search analysis genetic algorithm
\citep{geem2001}. 
Second, the uncertainties of the parameters are obtained via
simulated annealing around the best solution found by the genetic
algorithm. 

The free parameters are the scaled semi-major axis (\aRs, where
$a$ is the semi-major axis and \Rs\,is the stellar radius), the
planet-to-stellar radius ratio (\krprs, \Rp\, is the planetary
radius), the impact parameter $b$, where 
$b=a \cos i \sqrt{1-e^{2}} / \Rs (1 + e \sin \omega)$, $i$ is the
inclination, $e$ is the eccentricity, and $\omega$ is the argument of
the periastron. 
Values of $e$ and $\omega$ are known from the radial velocity
analysis, and their uncertainties were propagated allowing them to
vary between their $\pm 1\sigma$ uncertainties during the optimization
procedure.  
The epoch could also vary within the uncertainties reported in
Table~\ref{table:stellar_planetary_parameters}. 
Following \citet{csizmadia2013a}, limb darkening coefficients were
fitted as free parameters.
We took the results of \citet{brown2001} and \citet{pal2008a} into
account, fitting the combinations $u_{+} = u_{a} + u_{b}$ and  
$u_{-} = u_{a} - u_{b}$ rather than the linear and quadratic
coefficients of $u_{a}$ and $u_{b}$ individually.
A quadratic limb darkening law provided a reasonable fit to the data. 

\begin{figure*}
  \centering
  \begin{minipage}[t]{0.48\textwidth}
    \includegraphics[%
    width=0.9\linewidth,%
    height=0.5\textheight,%
    keepaspectratio]{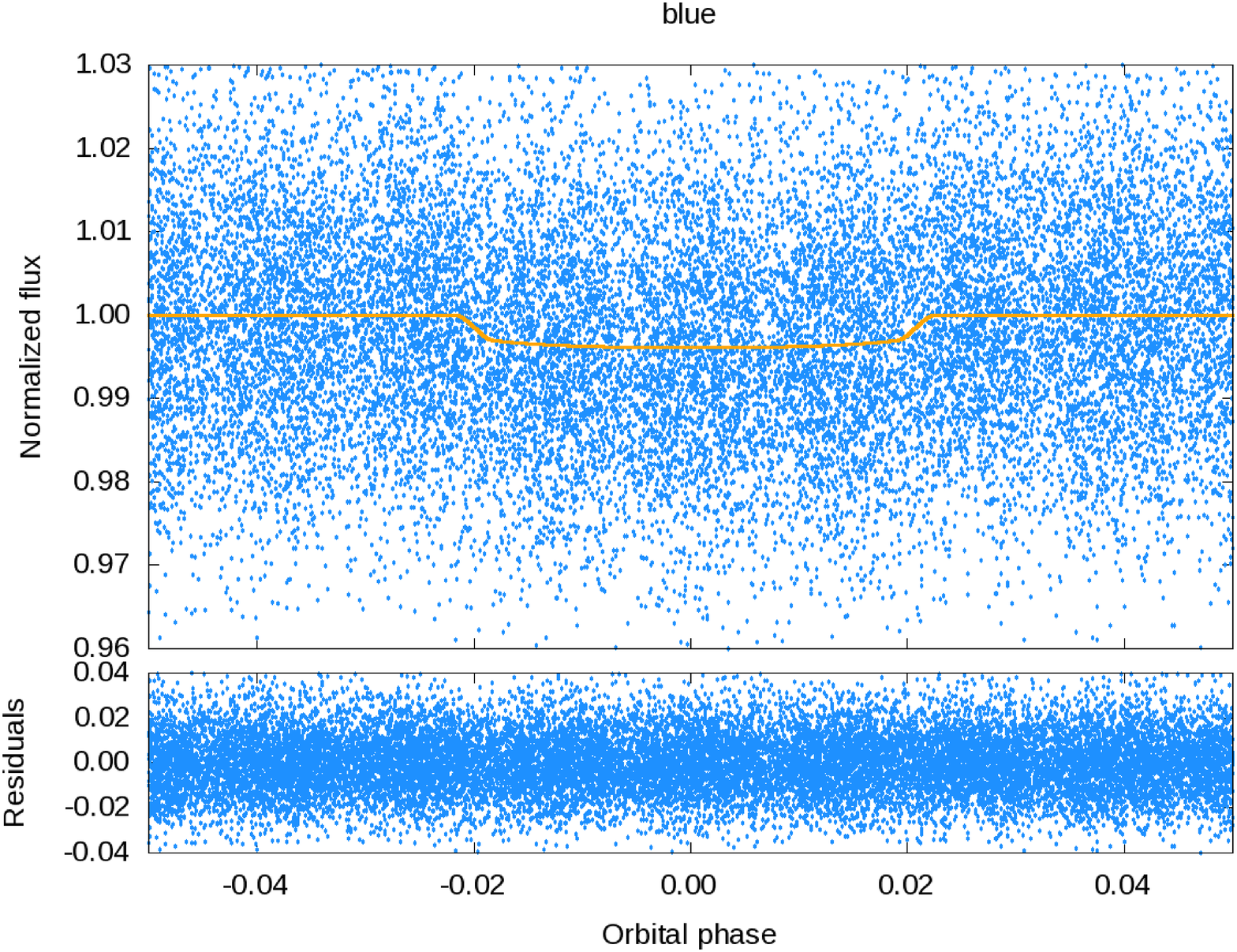}
    \includegraphics[%
    width=0.9\linewidth,%
    height=0.5\textheight,%
    keepaspectratio]{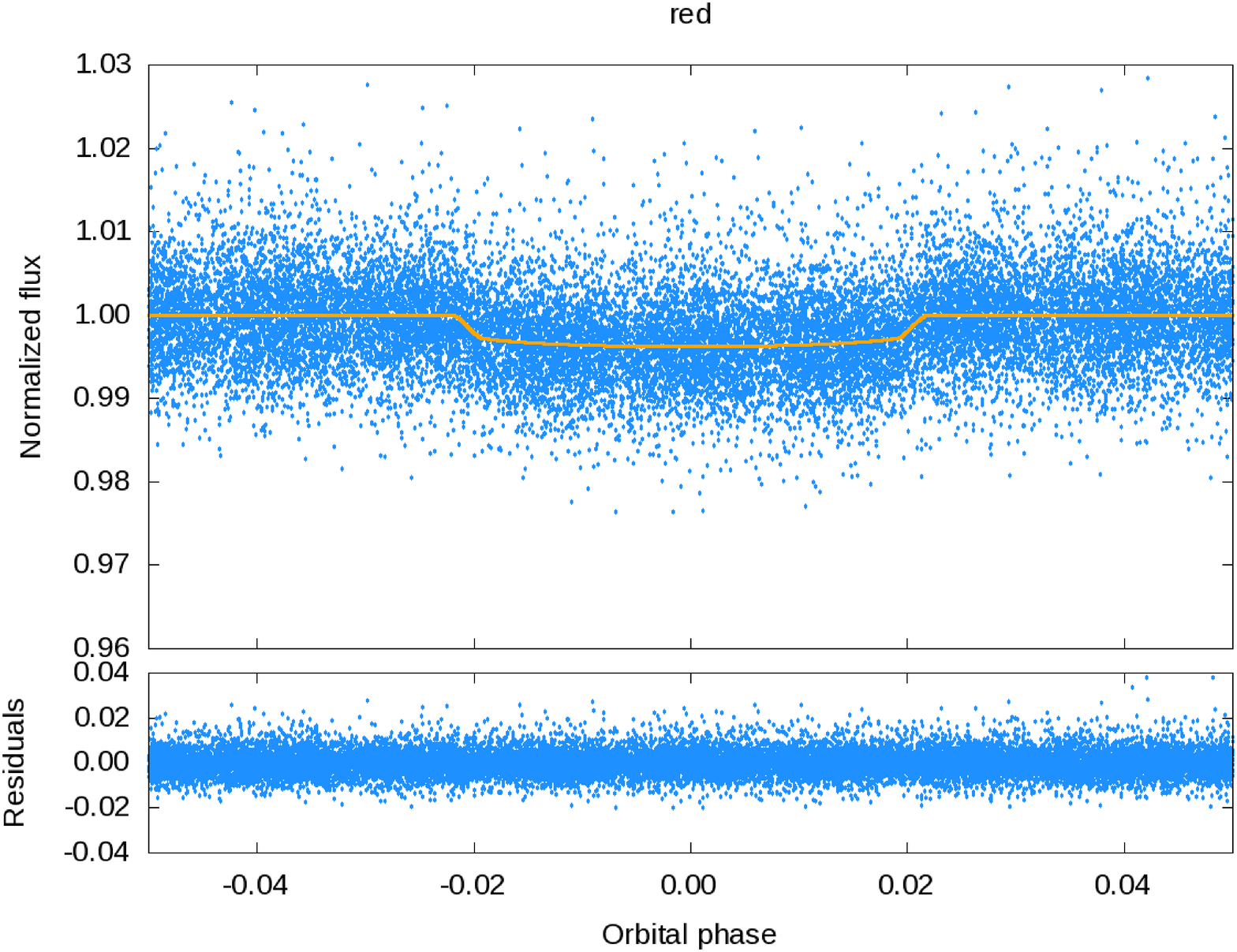}
  \end{minipage}
  \begin{minipage}[t]{0.48\textwidth}
    \includegraphics[%
    width=0.9\linewidth,%
    height=0.5\textheight,%
    keepaspectratio]{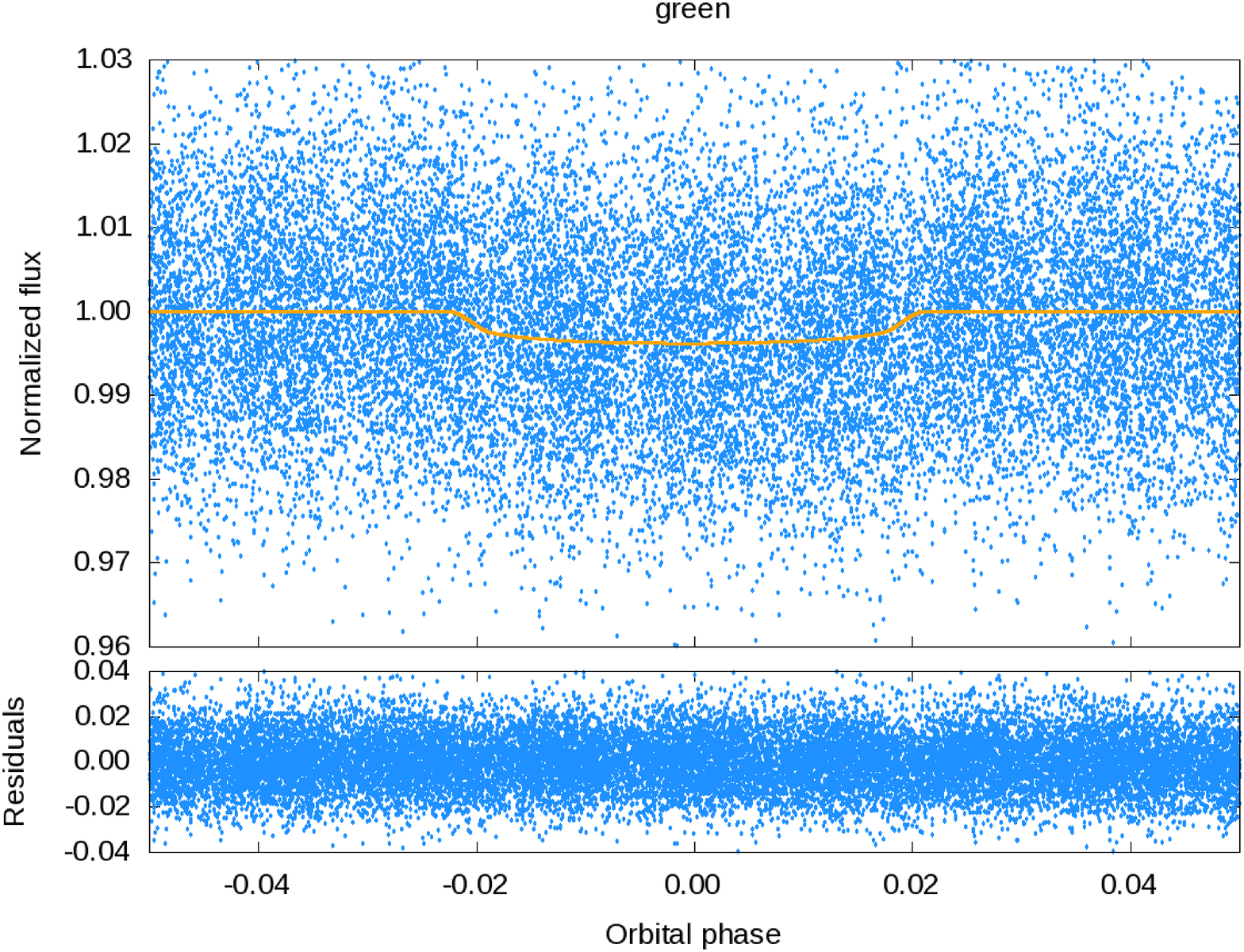}
    \includegraphics[%
    width=0.9\linewidth,%
    height=0.5\textheight,%
    keepaspectratio]{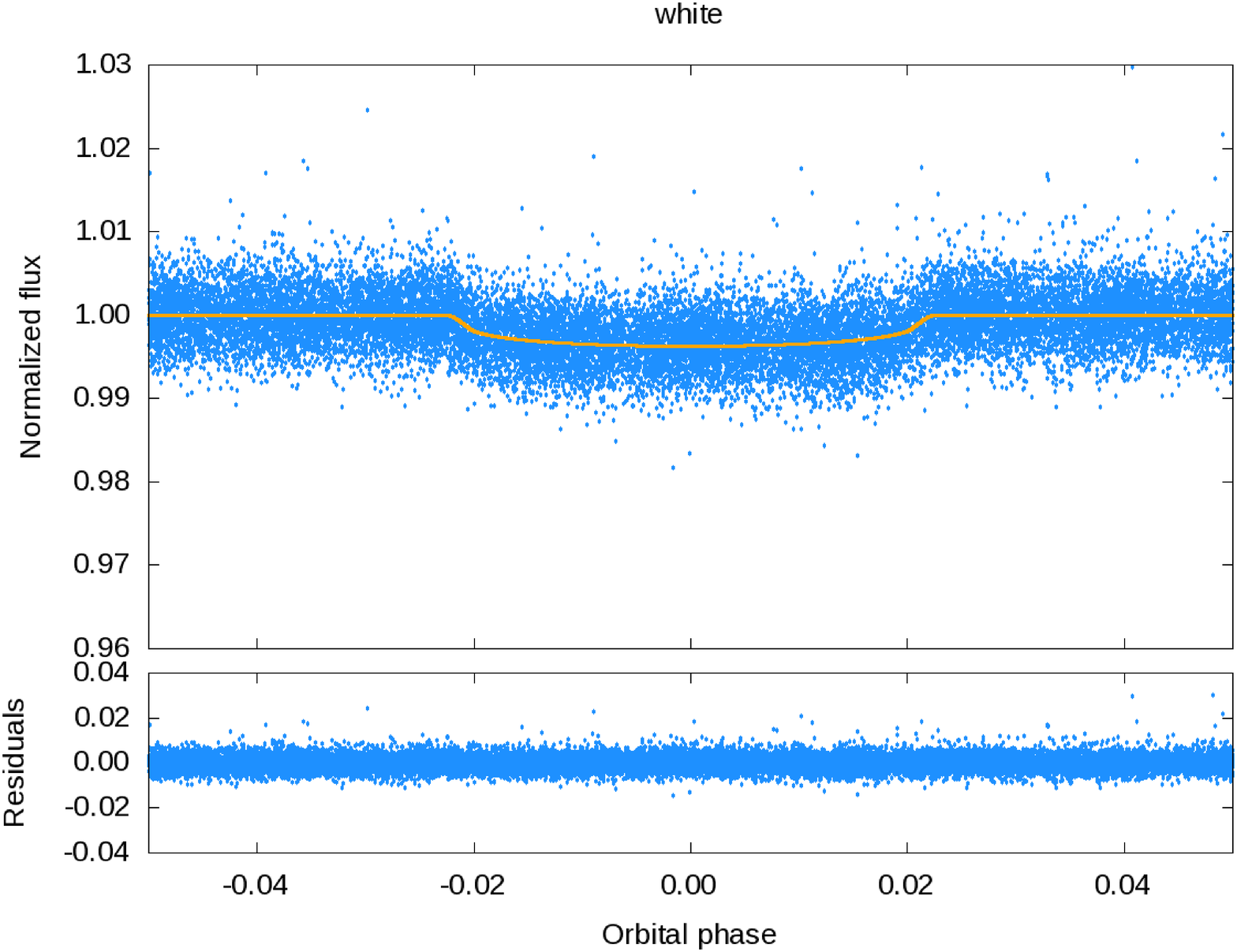}
  \end{minipage}
  \caption{The folded light curve of CoRoT-28b together with the best
    fit in the different chromatic light curves observed by \corot. 
    Blue points are the observed data; the orange curves represent the
    individual fits.}
  \label{figure:spherical_fit_c28b}
\end{figure*}

The fit is shown in Figure~\ref{figure:spherical_fit_c28b} and the
results can be seen in Table~\ref{table:spherical_fit_c28b}.
The fits obtained in different colours are in agreement with each
other.
Note that because of the non-axisymmetric chromatic PSF of \corot, the
contamination factors are not additive, so contamination in white is
not equal to the sum of contaminations in different colours. 
The white light curve results were used to establish the planetary
parameters given in Table~\ref{table:stellar_planetary_parameters}
because they have the smallest uncertainties because of  the higher SNR.

\begin{table*}
\caption{Parameters of the fit to the transit light curve of
  CoRoT-28b.} 
\label{table:spherical_fit_c28b}
\centering
\begin{tabular}{lcccc}       
\hline\hline 
Parameter          & \multicolumn{4}{c}{Value and uncertainty}                                                     \\ 
                   & in blue               &  in green             & in red                &  in white             \\
\hline
\aRs               & $    7.1 \pm 1.7    $ & $    6.7 \pm 2.7    $ & $   7.40 \pm 0.68   $ & $   7.29 \pm 0.16   $ \\
\krprs             & $ 0.0614 \pm 0.0073 $ & $ 0.0600 \pm 0.0107 $ & $ 0.0581 \pm 0.0033 $ & $ 0.0551 \pm 0.0004 $ \\
$b$                & $   0.43 \pm 0.36   $ & $   0.54 \pm 0.47   $ & $   0.30 \pm 0.26   $ & $   0.24 \pm 0.09   $ \\
$u_{+}$            & $   0.42 \pm 0.32   $ & $   0.71 \pm 0.40   $ & $   0.43 \pm 0.39   $ & $   0.78 \pm 0.11   $ \\
$u_{-}$            & $   0.32 \pm 0.62   $ & $   0.18 \pm 0.59   $ & $   0.40 \pm 0.58   $ & $   0.37 \pm 0.22   $ \\
contamination [\%] & $    6.2 \pm 3.0    $ & $    6.4 \pm 3.0    $ & $      2 \pm 1      $ & $      3 \pm 1      $ \\
$\redchisq$        & $   1.24            $ & $   1.28            $ & $   1.20            $ & $   1.32            $ \\
\hline
\end{tabular}
\end{table*}

\subsection{CoRoT-29b}
\label{subsection:planetary_parameters_c29b}

As described in Section~\ref{subsection:corot_reduction_corot29},
the light curve of CoRoT-29b has a non-flat, mid-transit shape.

The \corot\,data sets sampled at 512s and 32s cadence were treated
separately because of the background correction issue described in 
Section~\ref{subsection:corot_background_corot29}. 
The slope of the bottom part of the transit is non-zero with 95\%
confidence level in the 32s sampled data by \corot, and only non-zero
with 1$\sigma$ significance in the 512s sampled data set.
It is also non-zero with 99\% confidence level in the ground-based
data from FTN, and non-zero with 95\% confidence level in the IAC80
data (see Section~\ref{subsection:photometric_measurements}).
The slopes in the \corot\,data and in the ground-based data are
compatible with each other within 1$\sigma$ uncertainties, though the 
data from ground and from space were taken in slightly different
wavelengths and suffer from different kind of systematic effects. 

We have not been able to identify any systematic residual in the
\corot\,data, which could account for the asymmetry of the transit;
neither identified any other target in the \corot\,data set with a 
similar asymmetry.
Ground-based observations are affected by correlated noise, which can
produce asymmetries in the transit shape 
\citep[see, for example,][]{hebb2010,dittmann2010,mancini2013}.
But in this case, despite the inherent limitations of the ground-based
photometric observations that we present  (see 
Section~\ref{subsection:photometric_measurements}), we consider that
the asymmetry observed by \corot\,is confirmed by the ground-based
observations, and therefore it could have an astrophysical origin.

Possible interpretations for the asymmetry of the transit are given in
the next section (\ref{subsection:interpretations_c29b}), if the
2-sigma detection is assumed to be significant. 
The model that better reproduces the data is that  including the
gravity darkening produced by an oblate star (see
Section~\ref{subsubsection:interpretations_c29b_gravity_darkening}).  
We have taken those values as the reference planetary parameters 
(see Table~\ref{table:stellar_planetary_parameters}).

\subsection{Interpretations of the transit light curve of CoRoT-29b}
\label{subsection:interpretations_c29b}

\subsubsection{Stellar oblateness}
\label{subsubsection:interpretations_c29b_gravity_darkening}

We could interpret the shape of the light curve as the result of
the transit of the planet above a non-spherical stellar surface.  
A spherical surface becomes ellipsoidal, in a first approximation,
when it is deformed by its rotation, by its internal mass
distribution, or by the tidal influence of a massive companion. 
Consequently, the stellar poles are closer to the centre of the star
than the equator, thus the absorption rate of photons from the stellar
interior is different.
Additionally, the effective gravitational acceleration is higher at
the poles than at the equator, where the the centrifugal force reaches
its maximum. 
The decrease of the effective gravitational acceleration at the
equator affects the stellar atmospheric scale height, which in turn
changes the stellar temperature layering. 
All these effects cause a pole-to-equatorial variation in the stellar
flux, which is called gravity darkening, because this temperature
latitudinal variation is characterized by the effective surface
gravity $g$ at a certain latitude $b$ and with a gravity darkening
exponent $\beta$,  
\begin{equation}
T(b) = \Tpole \left( \frac{g(b)}{\gpole} \right)^\beta
.\end{equation}
The gravity darkening effect was studied by
\citet{zeipel1924,lucy1967,claret1999}, and \citet{claret2012} from
theoretical point of view. 
In the context of transiting extrasolar planets, this effect was
studied in depth by \citet{barnes2009a}.
Important observational results, in eclipsing binary stars and
transiting extrasolar planets, have been published by
\citet{rafert1980,nakamura1992,djurasevic2003,szabo2011,morris2013,barnes2013,zhou2013}.

As is shown hereafter, we successfully explained the observed transit
light-curve shape  of CoRoT-29b with the assumption that the asymmetry
of the light curve is caused by gravity darkening effect. 
We modelled this effect with our own code, which we briefly describe
next, and which will be describe in detail in a future publication
(Csizmadia, in prep.).
The optimization method is identical to the one described before 
(see Section~\ref{subsection:planetary_parameters_c28b}).

For this fit, we used the spectroscopic constraints for the \vsini\,
and \teff\,from the spectroscopic analysis of the star.
We allowed these parameters to vary between $3.5 \pm 0.5$ \kms\, and
$5260 \pm 70$K, respectively.
Here \teff\, is the surface temperature of the star averaged for the
visible hemisphere. 

The stellar shape is defined by $V=$constant equipotential
surfaces. 
We use the quadrupolar approach \citep[e.g.][]{zahn2010} instead of
the simple ellipsoidal approximation of \citet{barnes2009a}, 
\begin{equation}
\label{equation:potential}
V = -\frac{G\Ms}{R(b)} \left( 1 - J_{2} \left(\frac{\Rseq}{R(b)}\right)^{2} P_{2} (\sin b) \right) - \frac{1}{2} \OmegaRot^2 R^2(b)\cos^2 b
.\end{equation}
This equation can be rewritten in a dimensionless form as
\begin{equation}
\Psi = \frac{V \Rseq}{G\Ms} = -\frac{1}{r(b)} \left( 1 - \frac{J_{2}}{r(b)^2} P_{2} (\sin b) \right) - \frac{1}{2} \frac{w^2 r(b)^2 \cos^2 b }{(\aRseq)^3} 
.\end{equation}
Here we used the following notations:
\begin{itemize}
\item \Rseq, equatorial radius of the star,
\item $G$, gravitational constant,
\item \Ms, mass of the star,
\item $r(b) = R(b) / \Rseq$, the normalized radius of the star at
  latitude $b$,
\item \wPorbProt, a rotational parameter, namely the ratio of the
  orbital period of the planet and the rotational period of the star,
\item \aRseq, the scaled semi-major axis where $a$ is the semi-major
  axis of the planet,
\item $J_{2}$, the second (or quadrupole's) gravitational momentum of
  the star,
\item $P_{2}$, the second Legendre-polynomial.
\end{itemize}

\begin{figure}
  \centering
  \includegraphics[%
  width=0.9\linewidth,%
  height=0.5\textheight,%
  keepaspectratio]{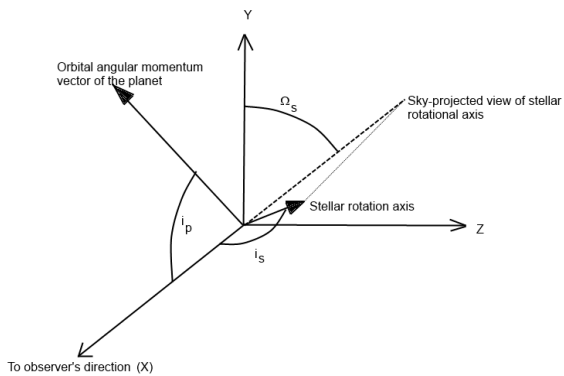}
  \caption{Angle convention used in the gravity darkening model.} 
  \label{figure:gravity_darkening_angle_convention}. 
\end{figure}

The free parameters are 
$J_{2}$; 
the gravity darkening exponent $\beta$;
the planet to stellar radius ratio $k=\Rp/\Rs$; 
the stellar inclination angle \Is (the obliquity of the star, as
defined in \citealt{barnes2009a}, is $\varphi = 90 - \Is$); 
\Omegas\,, the sky projected angle of the stellar spin axis ($\lambda$ in
the notation of \citealt{fabrycky2009a}); 
\Ip\,, the inclination of the planetary orbit; 
the limb darkening coefficients $u_{+}$ and $u_{-}$; 
the epoch of the transit $T_0$; 
the temperature of the star at the pole \Tpole; 
and the scaled semi-major axis \aRseq\,(see also
Figs.~\ref{figure:gravity_darkening_flow_chart} 
and~\ref{figure:gravity_darkening_angle_convention}). The rotational parameter $w$ was calculated iteratively from the known
orbital period and from \vsini, assuming a certain average radius of
the star.  
We chose a stellar radius of $1$ solar radius as initial value. 
After the first fit, we calculated the density parameter
$M^{1/3}/\Rseq = \left( 3 \pi \left( \aRseq \right)^3 / G \Porb^2 \right)^{1/3}$ for the star. 
We searched stellar models by \citet{hurley2000}, which provide this
density parameter at the effective surface temperature of the star
(within the uncertainties), and this yielded a new estimate for the
stellar radius. 
With this new stellar radius, we repeated the fit until convergence
(see Fig.~\ref{figure:gravity_darkening_flow_chart}).

The stellar parameters define the value of $\Psi$.
A Newton-Raphson iterative process yielded the value of $r$ and 
the actual value of the surface gravity $g=\nabla V$ for every
latitude and longitude. 
Then we calculated the effective surface temperature as well as the 
intensity assuming a simple black-body radiation for the star. 
A numerical integration procedure provided the unobscured flux of the 
star with a freely chosen gravity darkening exponent and limb
darkening coefficients.
As  is well known, the limb darkening coefficients are functions of
the effective stellar surface temperature.  
If the temperature varies over the surface, different limb darkening
coefficients will be valid at each surface point. 
Since the theoretical limb darkening tables are not always in
agreement with each other and they have not been validated
observationally, it is hard to accept that an interpolation of these
tables will provide the appropriate description of the limb darkening
coefficients for every surface point. 
In addition, a rotating stellar atmosphere can be different from a
static atmosphere, whereas most limb darkening tables use non-rotating
stellar atmospheric models.  
Therefore, we left the limb darkening coefficients free and used a
quadratic law hoping that this will be a better approximation of
reality in this particular case.  
For further details about the difficulties of limb darkening handling,
see \citet{csizmadia2013a}. 
Our code also includes the rotational beaming effect.

\begin{figure}
  \centering
  \includegraphics[%
  width=0.9\linewidth,%
  height=0.5\textheight,%
  keepaspectratio]{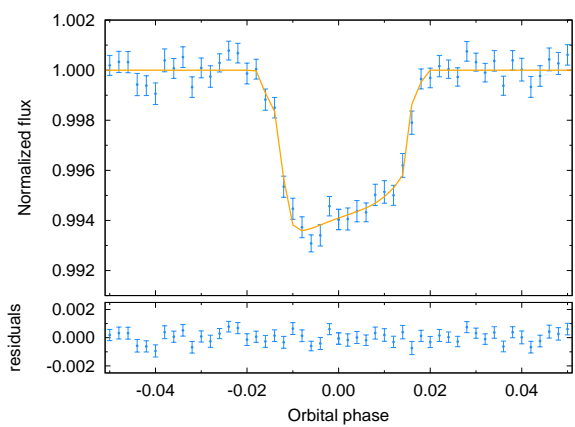}
  \caption{The binned and folded light curve of CoRoT-29b together
    with the best fit accounting for gravity darkening fit. 
    See text.} 
  \label{figure:gravity_darkening_fit_c29b}. 
\end{figure}

The planet's position was projected on the sky and when it crossed the
apparent stellar disc, we calculated the amount of the blocked stellar
emission in a small-planet approximation \citep{mandel2002}. 
Then we adjusted the free parameters until we had a convergence in the
\chisq-values, using the optimization methods described before 
(Section~\ref{subsection:planetary_parameters_c28b}).
The results of our modelling are shown in
Table~\ref{table:gravity_darkening_fit_c29b} and in graphical form in
Figure~\ref{figure:gravity_darkening_fit_c29b}. 
The final planetary parameters are also shown in
Table~\ref{table:stellar_planetary_parameters}.

\begin{table}
\caption{Parameters of the fit to the transit light curve of CoRoT-29b
  with a model accounting for gravity darkening.}
\label{table:gravity_darkening_fit_c29b}
\centering
\begin{tabular}{lc}       
\hline\hline 
\emph{Fitted parameters}            & Value and uncertainty \\                
\hline
$J_{2}$                             & $  0.028 \pm 0.019  $ \\
Gravity darkening exponent $\beta$  & $   0.89 \pm 0.25   $ \\
$k = \Rp/\Rs$                       & $ 0.1028 \pm 0.0043 $ \\
\Is\,[\degr]\tablefootmark{a}       & $     46 \pm 19     $ \\ 
\Omegas\,[\degr]\tablefootmark{b}   & $    256 \pm 40     $ \\
\Ip\,[\degr]                        & $   87.3 \pm 2.7    $ \\
$u_{+}$                             & $   0.62 \pm 0.15   $ \\
$u_{-}$                             & $   0.58 \pm 0.14   $ \\
$T_0$ [s]                           & $   -200 \pm 31     $ \\
\Tpole\,[K]                         & $ 5\,341 \pm 266    $ \\
\aRseq                              & $   9.22 \pm 0.19   $ \\
contamination factor [\%]           & $     54 \pm 4      $ \\
                                    &                       \\
\emph{Derived parameters}           &                       \\                
\hline
$f = \Rspo/\Rseq$                   & $   0.94 \pm 0.02   $ \\
\wPorbProt                          & $   0.26 \pm 0.06   $ \\
$\chisq$                            & $   71              $ \\
$\redchisq$\tablefootmark{c}        & $   1.4             $ \\ 
BIC                                 & $  120              $ \\
\end{tabular}
\tablefoot{
  \tablefoottext{a}{The obliquity of the star, as defined in \citet{barnes2009a}, is $\varphi = 90 - \Is = 47$\degr.}
  \tablefoottext{b}{The sky projected angle of the stellar spin axis; $\lambda$ in the notation of \citet{fabrycky2009a}.}
  \tablefoottext{c}{Sixty-two fitted points and 12 free parameters.}
}
\end{table}

\subsubsection{Spherical star model}
\label{subsubsection:interpretations_c29b_spherical}

\begin{figure}
  \centering
  \includegraphics[%
  width=0.9\linewidth,%
  height=0.5\textheight,%
  keepaspectratio]{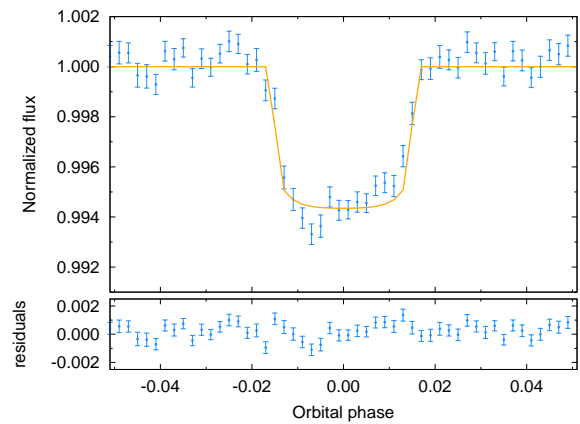}
  \caption{The binned and folded light curve of CoRoT-29b together
    with the best fit using a spherically symmetric model of the star.
    We discarded this model as a correct explanation of the measured
    light curve as explained in the text.}
  \label{figure:spherical_fit_c29b}
\end{figure}

If we decide to ignore the non-flat shape of the transit and try to
model the light curve of CoRoT-29b with the same techniques used in
the case of CoRoT-28b 
(see Section~\ref{subsection:planetary_parameters_c28b}), 
we obtain a fit solution significantly worse than using the more
sophisticated analysis including gravitational darkening.
The results are summarized in Table~\ref{table:spherical_fit_c29b} and 
in graphical form in Fig.~\ref{figure:spherical_fit_c29b}.
Notice the remarkable distribution of the binned points relative to
the fit. 
After the ingress, the observed points are systematically below the
fit, and later, during the mid-eclipse and just before the
egress phase they are systematically over the best match of the
spherical star fit. 
We conclude that we cannot ignore the asymmetry of the light curve in
the observed data set.

\begin{table}
\caption{Parameters of the fit to the transit light curve of CoRoT-29b
  using a spherically symmetric model for the star. 
  We discarded this model as a correct explanation of the measured
  light curve, as explained in the text.}
\label{table:spherical_fit_c29b}
\centering
\begin{tabular}{lc}       
\hline\hline 
Parameter                    & Value and uncertainty \\                
\hline
\aRs                         & $  10.49 \pm 0.72   $ \\
\Rp/\Rs                      & $ 0.0838 \pm 0.0070 $ \\
$b$                          & $   0.08 \pm 0.27   $ \\
$u_{+}$                      & $   0.89 \pm 0.33   $ \\
$u_{-}$                      & $  -0.49 \pm 0.38   $ \\
$\chisq$                     & $    103            $ \\
$\redchisq$\tablefootmark{a} & $   1.87            $ \\ 
BIC                          & $    132            $ \\
\hline
\end{tabular}
\tablefoot{
  \tablefoottext{a}{Sixty-two fitted points and seven free parameters.}
}
\end{table}

\subsubsection{Spot-occulting model}
\label{subsubsection:interpretations_c29b_spot}

Stellar spots occulted by transiting planets can have a significant
impact in the measured transit light curve. 
This phenomenon has been widely studied in the literature and has been
confirmed by several observations
(\citealt{silva2003}, \citealt{wolter2009} and references therein,
\citealt{sanchisojeda2012}, etc.).
We require from a spot-occulting solution to reproduce the observed
light curve with a stellar spot coverage, which is consistent with our
knowledge of the stellar properties.
Typically, the parameters used for this models are the stellar
temperature and rotational period and the spot temperature, size,
lifetime, and distribution.

In particular, the typical lifetime of a spot is $\sim 2$ weeks for
the Sun, but other stars have both shorter and longer lifetimes, up to 
several months in some cases \citep{queloz2001}.
In our case, comparing the identical \corot\,and the FTN observations
(see Section~\ref{subsection:photometric_measurements} and
Fig.~\ref{figure:ftn}), we require a spot lifetime larger than one year. 

Secondly, as all planetary transits observed by \corot\,in 2011
are indistinguishable with the current precision level, and they are
also indistinguishable from FTN data from 2012 and IAC80
ground-based observations from 2014, we must conclude that the
planet transits every 2.85 days above the same spot.
This requires an exact synchronization between the stellar rotational
period and the planetary orbital period, with exactly the same phase
in a baseline larger than three years.

On the one hand, if the observed asymmetry is related to an equatorial
or mid-latitude stellar spot, then the stellar rotational period
should be equal to the orbital period of the planet, otherwise we
cannot observe the spot always in phase with the planetary transits. 
However, we do not see a signature of such a spot as a modulation of
the light curve in phase with the stellar rotational period.
Moreover, we require the star to have a rotational period identical to
the orbital period, which is not consistent with the stellar
parameters derived from spectroscopy (in particular, with the \vsini). 

\begin{figure}
  \centering
  \includegraphics[%
  width=0.9\linewidth,%
  height=0.5\textheight,%
  keepaspectratio]{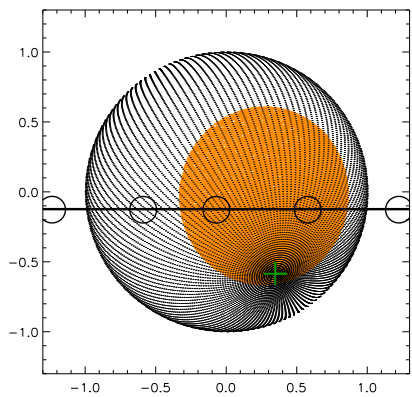}
  \caption{Visualization of the CoRoT-29 system according to the
    spot model. 
    The scale is in  stellar radii. 
    The orange area is the spotted area, while the horizontal line
    shows the sky-projected planetary orbit. 
    The positions of the planet at different orbital phases are shown;
    the size of the circles correspond to the planetary radius in
    this scale. 
    The pole of the star is denoted by a small green cross.}
  \label{figure:spot_model_c29b}
\end{figure}

On the other hand, if the spot is polar  and the rotational axis
of the star is perpendicular to the orbital plane of the planet, then
we could not observe an asymmetric light curve. 
A spot-caused asymmetry requires an off-position of the spot from the
apparent stellar disc. 
This scenario has the advantage that the signature of the spot is
diluted in the light-curve signal, and we could not infer from the
modulation of the light curve the stellar rotational period.
Additionally, if the spot is centred on the pole and the stellar
rotation axis has the right orientation, the planet could always cross
the same location of the spotted region. 
This requires no prior assumptions on the stellar rotation period (see
Fig.~\ref{figure:spot_model_c29b}). 

If we accept the polar spot scenario, however, which \emph{a priori}
cannot be ruled out as easily as the equatorial spot model, then we
must require that the polar spot does not change its apparent position
during the baseline of the observations.
In this model, the rotational axis of the star is very different from
the angular momentum of the vector of the planet, or in other words,
the planetary orbit is misaligned with the equatorial plane of the
star.   
The spot cannot be at the centre of the apparent stellar disc because
of the asymmetry of the light curve.
According to the light curve, the planet crosses the unspotted
surface of the star first, producing a deeper transit, and later the
spotted surface, when the lower brightness temperature of the spot
produces a shallower transit.
The apparent size of the spot should be much larger than the apparent
size of the planet because we observe a continuous change in the
transit depth and not a step-like event. 

Polar spots in active stars are a common phenomenon and in some cases
they can survive for long intervals of time.
Or at least, active regions continuously producing spots at similar
latitudes have been observed for several years in some young active
stars like V410 Tau \citep{hatzes1995,rice2011}.
But these polar spots are on very active, young stars that are rapidly
rotating (V410 Tau has a \vsini\,of 75\kms). 
Long-lived polar spots have never been seen, to our knowledge, on more
slowly rotating stars like CoRoT-29.
Moreover, these spots should produce signatures on the stellar spectrum
(\ion{Ca}{II} emission, distortion in the line profiles, 
flat-bottomed line profiles), which are not observed in our data set
(see Section~\ref{section:groundbased_observations}).

\begin{figure}
  \centering
  \includegraphics[%
  width=0.9\linewidth,%
  height=0.5\textheight,%
  keepaspectratio]{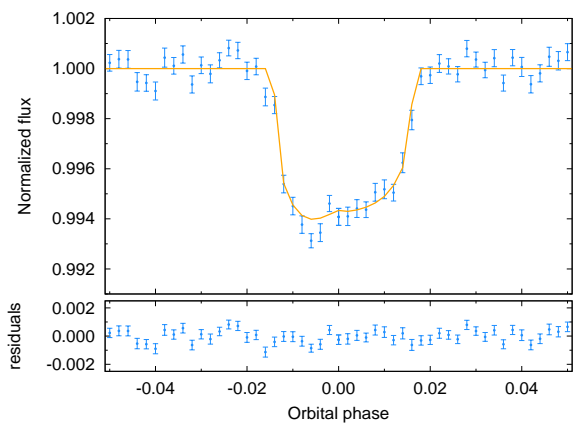}
  \caption{The binned and folded light curve of CoRoT-29b together
    with the red line being the best fit using a spherically symmetric
    model for the star and a stellar spot.
    For comparison, the solid green curve is because of  the unspotted
    light curve model (we set \Tspot = \Tstar in the fitting routine). 
    We discarded this model as a correct explanation of the measured
    light curve, as explained in the text.} 
  \label{figure:spot_fit_c29b}
\end{figure}

Although it is unlikely that a spot can survive for such a long period
without evolving on the surface of a main-sequence star, we used this
assumption in our modelling with a spherical star, a spherical planet,
and a circular spot on the surface of the star, with an arbitrarily
oriented stellar rotational axis and arbitrary initial spot
properties.  
We use a quadratic limb darkening law, as before.
The results are given in Table~\ref{table:spot_fit_c29b} and shown in
graphical form in Fig.~\ref{figure:spot_fit_c29b}. 

We obtain a polar spot with a tilted stellar rotational axis. 
This tilt means here that the projected rotational axis of the star 
and the projected angular momentum of the planet do not coincide. 
The planet is on 'pole-on' orbit because it orbits the star in such a 
way that its projected orbit crosses the polar region of the star.  
The mutual inclination between the stellar equator and planetary
orbital plane is practically 90 degrees. 
Note that such an orbit is very stable against perturbations.

From the measured \vsini\,and from the fitted stellar inclination for 
the spotted case (assuming $\Rs \sim 0.9\Rsun$), we find that 
the stellar rotational period should be between 0.7-8.4 days. 
The spectroscopically measured $\vsini \sim 3.5$ \kms\, would yield
$v_{eq} \sim 10$ \kms\, with $\Is \approx 20\degr$,
where the inclination of stellar rotational axis comes from the
spot fit. 

The comparison of the Bayesian Information Criterion (BIC) of the
spot fit and the gravity darkening fit makes a clear choice which
fit is better.
The spot fit yielded $BIC = 133,$ whereas the gravity darkening fit
produces $BIC = 120$. 
This BIC difference corresponds to a Bayes factor of around 665 for
the gravity darkening model (i.e. if the prior probabilities for
each model are the same, then the gravity darkening is 665 times
more probable than the spot model). 
Moreover, to explain the observational evidence we are obliged to put
requirements on the size and temporal evolution of the spot, which are
completely \emph{ad hoc}.
We consider that there is enough evidence to discard the spotted star
as a reasonable interpretation of the data.

An alternative to the single polar spot would be a band of spots
with short lifetimes, but continuously appearing during the baseline
of our observations in the same latitudes, relaxing one of the
requirements of the scenario. 
One possibility is that if the spots were conveniently appearing at 
different longitudes, the condition of the synchronous rotation
could also be relaxed.
Including more spots will improve the modelling results at the cost
of increasing the number of free parameters. 
Although plausible, it is finally also an \emph{ad hoc} solution for
the problem.
This hypothesis could  be falsified, however, if observations of the
Rossiter–McLaughlin effect would show a small spin-orbit angle.

\begin{table}
\caption{Parameters of the fit to the transit light curve of CoRoT-29b
  using a spherically symmetric model for the star and a stellar spot. The parameters  
  \bspot, \lspot\, and \dspot\, are the astrographic latitude,
  longitude, and diameter of the spot in degrees.  
  The stellar angles are chosen in the same way as in 
  Fig.~\ref{figure:gravity_darkening_angle_convention}.
  We discarded this model as a correct explanation of the measured
  light curve, as explained in the text.} 
\label{table:spot_fit_c29b}
\centering
\begin{tabular}{lc}       
\hline\hline 
Parameter            & Value and uncertainty \\                
\hline
\aRs                         & $  10.32 \pm 0.24   $ \\
\Rp/\Rs                      & $ 0.0948 \pm 0.0031 $ \\
\Ip                          & $   90.7 \pm 0.8    $ \\
$u_{+}$                      & $   0.90 \pm 0.19   $ \\
$u_{-}$                      & $  -0.26 \pm 0.28   $ \\
\Tstar\,[K] (fixed)          & $   5260            $ \\ 
\Tspot\,[K]                  & $   1600 \pm 400   $ \\
\bspot\,[\degr]              & $    -56 \pm 21     $ \\
\lspot\,[\degr]              & $    209 \pm 45     $ \\
\dspot\,[\degr]              & $     39 \pm 13     $ \\
\Is\,[\degr]                 & $    -42 \pm 20     $ \\
\Omegas\,[\degr]             & $    211 \pm 39     $ \\
contamination factor [\%]    & $     45 \pm 3      $ \\
$\chisq$                     & $     84            $ \\
$\redchisq$\tablefootmark{a} & $   1.75            $ \\
BIC                          & $    133            $ \\
\hline
\end{tabular}
\tablefoot{
  \tablefoottext{a}{Sixty fitted points and 12 free parameters.}
}
\end{table}

\subsubsection{Planetary oblateness}
\label{subsubsection:interpretations_c29b_oblate_planet}

A possible wind-driven oblated shaped of the planet can cause
asymmetries in the light curve, but the observable amplitude is about
100 ppm \citep{barnes2009b,zhu2014}, which is about 200 times smaller
than the observed effect in our case, therefore we reject this
explanation. 

Other examples of star-planet interaction do not provide
satisfactory answers in our case because they have typically 
different amplitudes and timescales
\citep[i.e.][]{jackson2012b,esteves2013,faigler2015}.

\subsubsection{Other scenarios}
\label{subsubsection:interpretations_c29b_crazy_ideas}

A disc around the star could produce an asymmetric transit, but there
are no signs of such disc in the light curve or in the infrared
emission of the star.
A moon around the planet could produce an asymmetric transit, but
dynamically there is no room for a moon around a planet so close to
its star.
A second planet in the system transiting simultaneously could produce
an asymmetric transit, but not every time that the transit is
observed, and there are no signs of this  planet in the radial
velocity.
We conclude that none of these scenarios provides a satisfactory
explanation of the data.

%
\begin{table*}
\caption{Planetary and stellar parameters.}            
\label{table:stellar_planetary_parameters}  
\begin{tabular}{l*{2}{r@{$\,\pm\,$}l}}
\hline\hline                 
                                                           & \multicolumn{2}{c}{CoRoT-28b}     & \multicolumn{2}{c}{CoRoT-29b}       \\
\multicolumn{5}{l}{\emph{Ephemeris}}                                                                                                 \\
\hline
Planet orbital period $P$ [days]                           & $     5.208\,51 $ & $ 0.000\,38 $ & $  2.850\,570$ & $ 0.000\,006 $ \\ 
Transit epoch $T_{\mathrm{tr}}$ [HJD-2\,450\,000]          & $ 5\,755.846\,6 $ & $ 0.002\,9  $ & $ 5\,753.115 $ & $ 0.001      $ \\
Transit duration $d{_\mathrm{tr}}$ [h]                     & $          5.70 $ & $ 0.09      $ & $       2.19 $ & $ 0.04       $ \\
\multicolumn{5}{l}{\emph{Results from radial velocity observations}} \\
\hline    
Orbital eccentricity $e$                      & \multicolumn{2}{c}{$ 0.047^{+0.055}_{-0.038}  $} &                    $     0.082 $ & $ 0.081    $  \\
Argument of periastron $\omega$ [\degr]       & \multicolumn{2}{c}{$ 126^{+140}_{-61}         $} & \multicolumn{2}{c}{$ 87^{+210}_{-49}          $} \\ 
Radial velocity semi-amplitude $K$ [\ms]      &                    $      56.4 $ & $ 4.9      $  &                    $       125 $ & $ 17       $  \\
Systemic velocity $V_{r}$ (HARPS) [\kms]      &                    $   76.7081 $ & $ 0.0051   $  &                    $   64.1907 $ & $ 0.0012   $  \\
O-C residuals (HARPS) [\ms]                   & \multicolumn{2}{c}{$ 6.4^{+7.0}_{-4.8}        $} &                    $        11 $ & $ 11       $  \\
Systemic velocity $V_{r}$ (Sophie) [\kms]     &                    $   76.7549 $ & $ 0.0082   $  & \multicolumn{2}{c}{$                          $} \\
O-C residuals (Sophie) [\ms]                  &                    $      32.8 $ & $ 6.4      $  & \multicolumn{2}{c}{$                          $} \\
Systemic velocity $V_{r}$ (FIES) [\kms]       &                    $   76.8642 $ & $ 0.0074   $  & \multicolumn{2}{c}{$                          $} \\
O-C residuals (FIES) [\ms]                    &                    $       8.5 $ & $ 8.3      $  & \multicolumn{2}{c}{$                          $} \\
\multicolumn{5}{l}{\emph{Fitted transit parameters}} \\
\hline
Scaled semi-major axis \aRseq                                  &                    $  7.29   $  & $ 0.16    $  &                    $  9.22   $ & $  0.19   $  \\
Radius ratio \krprs                                            &                    $  0.0551 $  & $ 0.0004  $  &                    $  0.1028 $ & $  0.0043 $  \\
Quadratic limb darkening coefficients\tablefootmark{a} $u_{+}$ &                    $  0.78   $  & $ 0.11    $  &                    $  0.62   $ & $  0.15   $  \\
\phantom{Quadratic limb darkening coefficients\,}      $u_{-}$ &                    $  0.37   $  & $ 0.22    $  &                    $  0.58   $ & $  0.14   $  \\
Impact parameter\tablefootmark{b} $b$                          &                    $  0.24   $  & $ 0.09    $  & \multicolumn{2}{c}{$-                      $} \\
Orbital inclination \Ip\,[deg]                                 & \multicolumn{2}{c}{$-                       $} &                    $ 87.3    $ & $  2.7    $  \\
Stellar rotational axis inclination \Is\,[\degr]               & \multicolumn{2}{c}{$-                       $} &                    $ 46      $ & $ 19      $  \\
Stellar rotational axis orientation \Omegas\,[\degr]           & \multicolumn{2}{c}{$-                       $} &                    $ 256     $ & $ 40      $  \\
Gravity darkening exponent $\beta$                             & \multicolumn{2}{c}{$-                       $} &                    $ 0.89    $ & $ 0.25    $  \\
$J_{2}$ for the star                                           & \multicolumn{2}{c}{$-                       $} &                    $ 0.028   $ & $ 0.019   $  \\
Contamination factor [\%]                                      &                    $  3      $  & $ 1       $  &                    $ 54      $ & $ 4       $  \\
\multicolumn{3}{l}{\emph{Deduced transit parameters}} \\
\hline
Stellar radius ratio $f = \Rspo/\Rseq$             & \multicolumn{2}{c}{ 1 (fixed)         } &                    $ 0.94   $ & $ 0.02 $  \\
$\Ms^{1/3}/\Rs$ [solar units]\tablefootmark{c}     &                    $ 0.58 $ & $ 0.02 $  &                    $ 1.09   $ & $ 0.02 $  \\
Stellar density $\rho_{s}$ [\kgm3]                 &                    $ 270  $ & $ 6    $  &                    $ 1\,825 $ & $ 38   $  \\ 
Impact parameter\tablefootmark{b} $b$              & \multicolumn{2}{c}{$-                $} &                    $ 0.43   $ & $ 0.44 $  \\ 
Orbital inclination \Ip\,[deg]                     &                    $ 88.1 $ & $ 0.8  $  & \multicolumn{2}{c}{$-                  $} \\

\multicolumn{3}{l}{\emph{Spectroscopic parameters }} \\
\hline
Effective temperature \teff\,[K]                     &                    $ 5\,150 $ & $ 100  $  &                    $ 5\,260 $ & $ 100  $  \\
Stellar surface gravity \logg\,[cgs]                 &                    $    3.6 $ & $ 0.2  $  &                    $    4.3 $ & $ 0.2  $  \\
Metallicity $[\rm{Fe/H}]$ [dex]                      &                    $   0.15 $ & $ 0.10 $  &                    $   0.20 $ & $ 0.10 $  \\
Stellar rotational velocity {\vsini} [\kms]          &                    $    3.0 $ & $ 0.5  $  &                    $    3.5 $ & $ 0.5  $  \\
Microturbulence velocity $V_{\mathrm{mic}}$ [\kms]   &                    $    0.9 $ & $ 0.1  $  &                    $    0.9 $ & $ 0.1  $  \\
Macroturbulence velocity $V_{\mathrm{mac}}$ [\kms]   &                    $    3.3 $ & $ 0.3  $  &                    $    1.2 $ & $ 0.3  $  \\
Lithium equivalent width [m\angstrom]                & \multicolumn{2}{c}{$ 31                $} & \multicolumn{2}{c}{ no Li at 6708\AA    } \\
Spectral type                                        & \multicolumn{2}{c}{ G8/9IV              } & \multicolumn{2}{c}{ K0~V                } \\
\multicolumn{2}{l}{\emph{Stellar and planetary physical parameters from combined analysis}} \\
\hline
Star mass [\Msun]\tablefootmark{c}                                &                    $   1.01 $ & $ 0.14   $  &                    $   0.97 $ & $ 0.14   $  \\
Star radius [\Rsun]\tablefootmark{c}                              &                    $   1.78 $ & $ 0.11   $  &                    $   0.90 $ & $ 0.12   $  \\
Stellar surface gravity \logg\,[cgs]                              &                    $   3.94 $ & $ 0.12   $  &                    $   4.52 $ & $ 0.19   $  \\
Age of the star $t$ [Gyr]                                         &                    $   12.0 $ & $ 1.5    $  & \multicolumn{2}{c}{$        1 - 8        $} \\
Distance of the system [pc]                                       &                    $    560 $ & $ 70     $  &                    $    765 $ & $ 50     $  \\
Interstellar extinction \Av\,[mag]                                &                    $   0.75 $ & $ 0.20   $  &                    $   0.85 $ & $ 0.15   $  \\
Orbital semi-major axis (Kepler's law) $a$ [AU]                   &                    $ 0.0590 $ & $ 0.0027 $  &                    $ 0.0390 $ & $ 0.0020 $  \\
Orbital semi-major axis (light curve) $a$ [AU]                    &                    $ 0.0603 $ & $ 0.0050 $  &                    $ 0.0386 $ & $ 0.0059 $  \\
Planet mass \Mp\,[\Mjup]\tablefootmark{d}                         &                    $  0.484 $ & $ 0.087  $  &                    $   0.85 $ & $ 0.20   $  \\
Planet radius \Rp\,[\Rjup]\tablefootmark{d}                       &                    $  0.955 $ & $ 0.066  $  &                    $   0.90 $ & $ 0.16   $  \\
Planet density $\rho_{\mathrm{p}}$ [\gcm3]                        &                    $   0.69 $ & $ 0.27   $  &                    $   1.45 $ & $ 0.74   $  \\
Planet surface gravity log\,$g$ [cgs]                             &                    $   3.12 $ & $ 0.14   $  &                    $   3.42 $ & $ 0.19   $  \\
\hline       
\end{tabular}
\tablefoot{
  \tablefoottext{a}{$I(\mu)/I(1)= 1-\mu+u_{a}\mu + u_{b} (1-\mu)^2$,
  where $I(1)$ is the specific intensity at the centre of the disc and 
  $\mu=\cos{\gamma}$, $\gamma$ being the angle between the surface
  normal and the line of sight; $u_{+}=u_{a}+u_{b}$ and
  $u_{-}=u_{a}-u_{b}$.}
  \tablefoottext{b}{$b=a \cos i \sqrt{1-e^{2}} / \Rs (1 + e \sin \omega)$}
  \tablefoottext{c}{Radius and mass of Sun taken as $695\,500$ km
  and $1.9891 \times 10^{30}$ kg, respectively \citep{lang1999}.}
  \tablefoottext{d}{Radius and mass of Jupiter taken as $71\,492$ km
  and $1.8992 \times 10^{27}$ kg, respectively \citep{lang1999}.}
}
\end{table*}

%
\section{Discussion}
\label{section:discussion}

\subsection{Stellar properties}
\label{subsection:stellar_properties}

\noindent {\it CoRoT-28}\\\mbox{ }

\begin{figure}
  \centering
  \includegraphics[%
  width=0.9\linewidth,%
  height=0.5\textheight,%
  keepaspectratio]{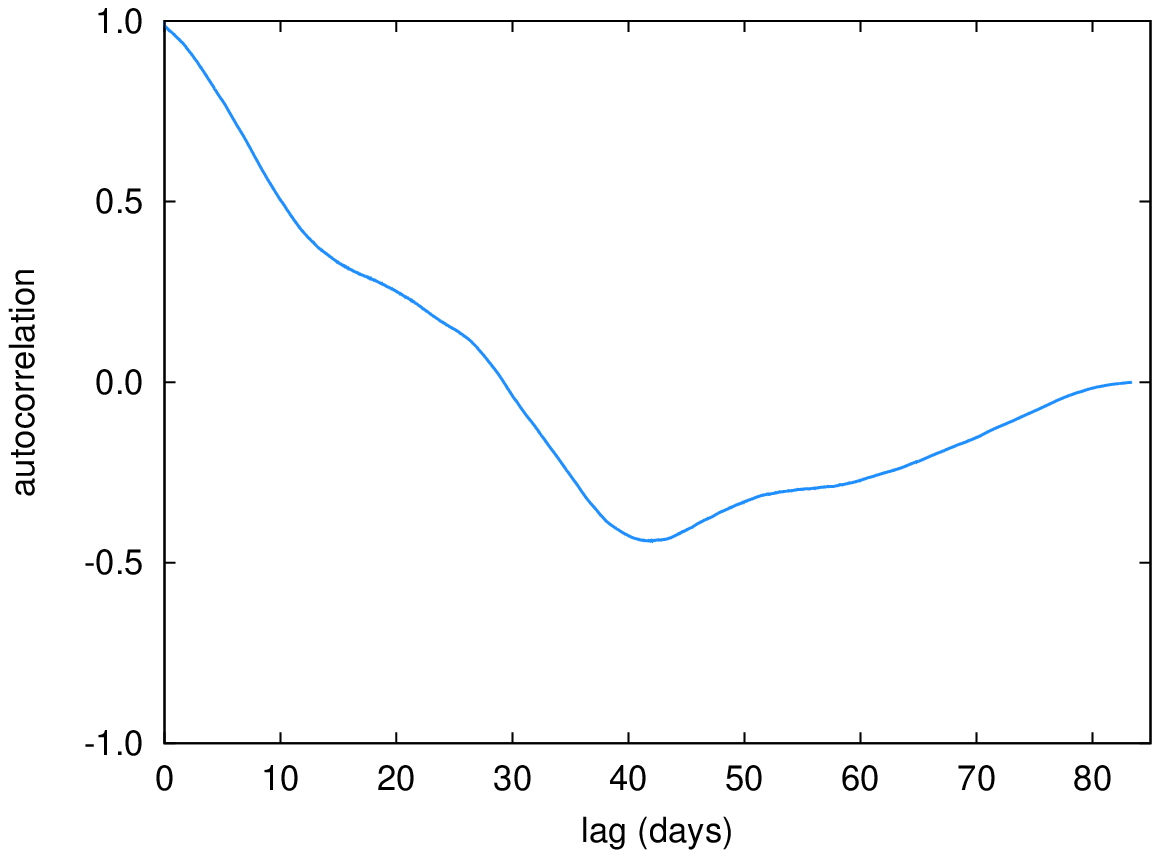}
  \caption{Autocorrelation function of the light curve of CoRoT-28.}
  \label{figure:corot28_acf}
\end{figure}

CoRoT-28 is a G8/9IV evolved star with a radius of approximately 1.8 
solar radii.  
With an age of approximately 12 Gyr, we expect from gyrochronology a
rotational period around 70 or 80 days and a \vsini\,around
1.2\kms\,or smaller \citep{barnes2007}.
The \vsini\,value from spectroscopy is almost three times larger, giving
an expected rotational period of about 30 days.
The autocorrelation function (ACF) of the light curve 
\citep[see][]{mcquillan2013a} does not show any evident periodicity at
all, but if there is any sign, it is for periods larger than the 85
days observed by \corot (see Fig.~\ref{figure:corot28_acf}). 

There is an inconsistency between the \vsini\,determined by
spectroscopy, which is consistent with a relatively fast rotational
period, and the stellar parameters derived from the models, which
suggest an old star with a slow rotation rate.
From our analysis of the light curve we cannot decide between the
two scenarios. 
The \vsini\,has a relatively large uncertainty, but seems to be
consistent with the higher rotation rate.
It is possible that the stellar models fail to reproduce the actual
age of the star. 
The range proposed by the models is $12.0 \pm 1.5$ Gyr, close to the
old end of the distribution of stellar populations, which is
consistent with the evolved stellar radius from the models 
($1.78 \pm 0.11$ \Rs\,for an effective temperature of $5\,150 \pm 100$
K). 
This range, however, is perhaps not as consistent with the metallicity 
$[\rm{Fe/H}] = 0.15 \pm 0.05$ dex, which is  quite a robust
observational result. 

There is also an apparent inconsistency between the age and the
\ion{Li}{I} abundance in the atmosphere of the star.
We estimated $\log N(Li)$ using Table~2 of \citet{soderblom1993} given
the equivalent width of 31m\angstrom. 
For a dwarf with an effective temperature of 5\,150K, we obtained
$\log N(Li)=1.4$. 
This is in line with an age of a few hundred Myrs according to
\citet{sestito2005}. 
However, stars of this effective temperature with ages of few hundred
Myrs have already reached the main sequence, which is not compatible
with the surface gravity measured from spectroscopy or with the
transit duration, which indicates that the star has a radius larger
than the Sun.
Moreover, we would expect a significant degree of stellar variability
and a faster rotation period, which is also not supported by the
observations.
The abundance of lithium and its relationship with age, especially for
stars hosting planets 
\citep[see, for example, the recent discussions by][]{liu2014,adamow2012}, 
is still debated.
Therefore we do not consider lithium as a reliable age indicator for
CoRoT-28.  
On the other hand, it is known that a small amount of \ion{Li}{I} can
also be present in evolved stars, for example lithium can be
regenerated by the Cameron-Fowler mechanism \citep{cameron1971}.

We consider that observational evidence speaks in the case of
CoRoT-28  in favour of an evolved star with a higher lithium content
and \vsini\, value than what would be expected for its age and
evolutionary status.

\mbox{ }\\\noindent {\it CoRoT-29}\\\mbox{ }

\begin{figure}
  \centering
  \includegraphics[%
  width=0.9\linewidth,%
  height=0.5\textheight,%
  keepaspectratio]{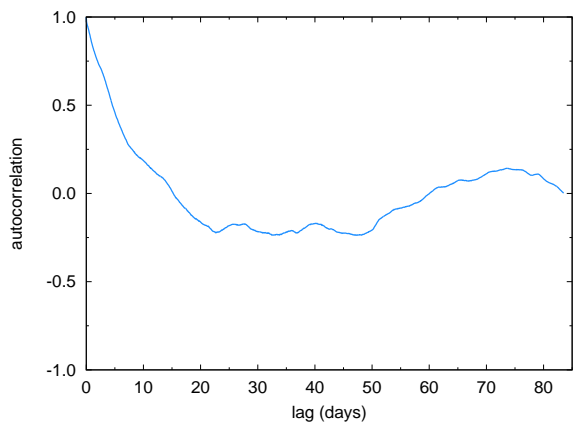}
  \caption{Autocorrelation function of the light curve of CoRoT-29.}
  \label{figure:corot29_acf}
\end{figure}

CoRoT-29 is a K0 main-sequence star with a radius of approximately 0.9
solar radii.
Its age is not well constrained from the analysis of the stellar
models and the autocorrelation function of the light curve (see 
Fig.~\ref{figure:corot29_acf}) does not help to constrain the
rotational period. 
In any case, the analysis of the ACF in the case of CoRoT-29 can be
debated, as $\sim 50\%$ of the flux within the mask comes from a
contaminant. 

It is however interesting that the analysis of the light curve
(Section~\ref{subsection:interpretations_c29b}) shows that 
the orbital plane of the planet could be misaligned with respect
to the equatorial plane of the star. 
The gravity darkening model allows us to measure the relative 
inclination of the spin axis of the star, a parameter that typically
cannot be constrained with photometric measurements. 

The alignment between the vector normal to the planetary orbital plane
and the stellar spin axis is considered one of the key observables
constraining planetary formation and evolution 
\citep[for a recent discussion, see][and references therein]{dawson2014}.
It has been proposed that the alignment, or misalignment, is related
to the dynamical interaction between the star and the planet, which in
turn depends on the internal distribution of matter inside the star.
Cold, convective stars have different dissipation timescales than hot,
radiative stars \citep[see][]{albrecht2012}. 
CoRoT-29 with a \teff\,of $5\,260$K and with a misalignment 
angle\footnote{\Omegas\,is the sky projected angle of the stellar spin
  axis, $\lambda$ in the notation of \citet{fabrycky2009a}. In the
  notation of \citet{albrecht2012}, it would be $\lambda = 180 -
  \Omegas$.} of $\Omegas = 301 \pm 30$\degr would be right between the
outliers of the distribution HAT-P-11 and HD 80606 in Figure~20 of 
\citet{albrecht2012}.
It is however worth mentioning that the tidal dissipation timescale of
CoRoT-29b is 1000 times faster than for HAT-P-11, using equation 2
from the same paper. 
As mentioned, the dissipation timescale strongly depends on the
internal structure of the star, which deserves further attention in
the case of CoRoT-29. 

\begin{figure}
  \centering
  \includegraphics[%
  width=0.9\linewidth,%
  height=0.5\textheight,%
  keepaspectratio]{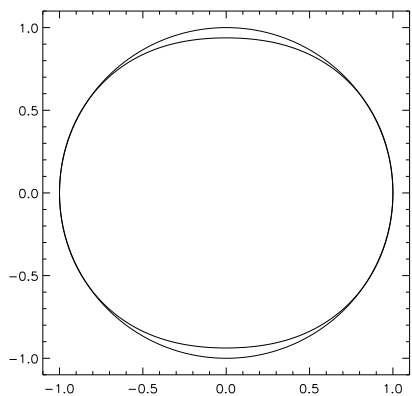}
  \caption{Shape of the stellar surface of CoRoT-29 in the gravity
    darkening model (thick line) compared with a spherically symmetric
    star (thin line).}
  \label{figure:stellar_shape_c29}
\end{figure}

The shape of the transit light curve of CoRoT-29b has a strong
contribution from the gravity darkening, orders of magnitude larger
than in other examples found \citep[i.e.][]{szabo2011,zhou2013}. 
The contributions are only comparable in the case of the pre-main-sequence star PTFO~8-8695 \citep{barnes2013}. 
However, the origin of the gravity darkening effect in PTFO~8-8695
is the fast rotation of the star, a M dwarf with a radius of 1.4\Rsun,
a rotational period of 10 hours, and a \vsini\,of 80\kms. 
For CoRoT-29b, the origin of the extreme pole to equatorial radius
difference is not as easily understood 
(see Fig.~\ref{figure:stellar_shape_c29}).
The spectroscopically measured \vsini\, and the resulted
stellar inclination ($43$\degr) yield $v_{\mathrm{eq}} = 5.1 \pm
1.8$ \kms\, and that corresponds to a stellar rotational period of
4-13 days.

To lowest order, and in the assumption of uniform rotation, the
stellar shape is defined by the surface of constant total potential as
defined in Eq.~\ref{equation:potential}. 
It shows the balance between the contribution to the local
gravitational potential of the stellar rotation \OmegaRot\, and the
quadrupole moment term dominated by $J_{2}$. 
Actually, we should have included in the balance the contribution from
the tidal distortion caused by the planet, and account for the
misalignment between the rotation axis of the star and the angular
momentum of the orbit. 
But when comparing the tidal and rotational contributions to the
distortion (see equation 3 of \citealt{ragozzine2009} or A.21 of
\citealt{leconte2011}), it turns out to be a couple orders of magnitude
less important, and so we neglect it.
Note also that in equation~\ref{equation:potential} the contribution
from rotation is negligible compared to the contribution of the
$J_{2}$ term.
This means that in our modelling the rotation rate is constrained
mainly by the \vsini\,value. 
The model can measure the deformation and the orientation of the star,
but not its rotation rate.
The results of Section~\ref{subsection:planetary_parameters_c29b} show
that for CoRoT-29, the quadrupole moment has a contribution 2 orders of
magnitude larger than rotation.
Actually, the quadrupole moment of CoRoT-29 measured from the gravity
darkening is far too large, $J_{2} = 0.028 \pm 0.019$, compared to the
Sun, $J_{2}^{\odot} = \left( 1.7 \pm 0.4 \right) \cdot 10^{-7}$
\citep{lang1999}. 
Other authors have claimed large $J_{2}$ values for fast-rotating
stars, such as WASP-33 \citep[$J_{2} = 3.8 \cdot
10^{-4}$,][]{iorio2011}, but not as large as for CoRoT-29.

The value of the quadrupolar moment depends on the distribution of
matter inside the star. 
When the star is in uniform rotation, this quadrupolar moment is
closely related to the Love number\footnote{Note that $k_{2}$ is twice
  the apsidal motion constant, often indicated with the same symbol as
  in, e.g. \citet{claret1995}.} $k_{2}$ in the linear approximation. 
Indeed, once $k_{2}$ is known, the external potential and the shape
that a body  exhibits in response to any perturbing potential can
be computed \citep[see, for example,][]{leconte2011}. 
The Love number is a measure for the level of central condensation of
an object: $k_{2}$ decreases as the degree of central condensation
increases. 
A homogeneous incompressible ideal fluid body has $k_{2}=3/2$. 
For a star like CoRoT-29, the value of $k_{2}$ can be estimated from
canonical stellar models \citep[see, for example,][]{claret1995}. 
Considering the \vsini\,value and the mass and orbital distance of the
companion, the expected value of $k_{2}$ for CoRoT-29 would produce a
modest distortion, resulting in a $J_{2}$ that would be 3 to 4 orders
of magnitude smaller than our estimated value. 
In other words, the star is not rotating fast enough and the planetary
companion is neither massive enough nor close enough to significantly
distort the stellar shape. 
However, it is expected that the flattening may assume high values
when the star is rotating differentially, with the angular velocity
decreasing outwards \citep{jackson2004,zahn2010}. 
In turn, in the assumption of shellular rotation, a stellar rotation
regime where the angular velocity is constant on level surfaces but
varying in depth, \citep{zahn2010} has shown that $J_{2}$ also
increases compared to the case of uniform rotation. 
Assuming an internal differential rotation with a centre-to-surface
ratio of 4 in angular velocity, the $J_{2}$ of a 7\Msun\,rapidly
rotating star like Archenar can exceed 0.01, which is about twice the
value it assumes under the hypothesis of uniform rotation. 
It is not straightforward to infer how such a result could be
extrapolated to a much less massive star and with slow surface
rotation such as CoRoT-29. 
The star seems to be distorted and it is apparently a
robust result, observed from space and independently confirmed from
ground. 
The gravity darkening model provides the most accurate and reasonable 
explanation, although a dedicated modelling out of the scope of this
paper would be required. 

\subsection{Star-planet interaction}
\label{subsection:star-planet_interaction}

In the previous section, we have mentioned that the stars seem to
rotate faster than expected for their evolutionary stage.
In the case of CoRoT-29, for example, if magnetic braking is acting in
this K0V star, and it is expected to be, the rotation should be
slower. 
So, either the system is very young, or we have to find a mechanism to
speed up the stellar rotation
One way to speed up rotation is tidal interaction between the star and
the planet.
But the planets CoRoT-28b and CoRoT-29b may not be very efficient for
that reason because of their orbital properties and, in the case of
CoRoT-29b, for the high inclination of its orbit over the stellar
equator  \citep[for a review of star-planet interaction in the
\corot\,context, see][]{paetzold2012}.

One hypothesis is that a second planet was present that accelerated
the stellar rotation before falling into the star. 
Under this assumption, two planets formed simultaneously far away from
the star (from the current position of the transiting planet).
Planet-planet scattering could have moved then one planet in an inner
orbit, very close to the star, and an outer planet in a larger,
eccentric orbit.
Tidal interactions would have made the inner planet fall into the
star, accelerating its rotation rate and enriching its atmosphere with
heavier elements such as lithium, while the outer planet would have
subsequently circularized its orbit until the conditions that we see
today. 

To assess the viability of this kind of a scenario, we investigated
the behaviour of such a hypothetical planet in the neighbourhood of
the star. 
Although it is impossible to find observational evidence for this
former planet, it is still useful to explore plausible mechanisms that
can explain the current properties of CoRoT-29.

\begin{figure}
  \centering
  \includegraphics[%
  width=0.9\linewidth,%
  height=0.8\textheight,%
  keepaspectratio]{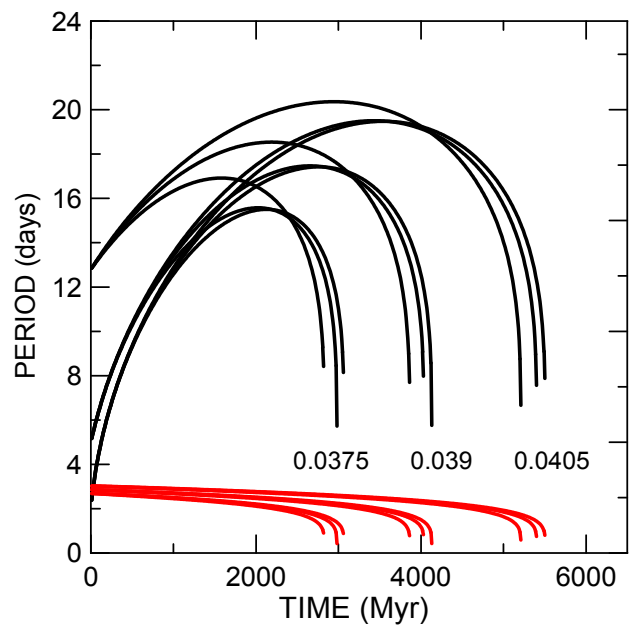}
  \caption{Evolution of the period of rotation of the star in days.
    The red curve is the orbital period of the hypothetical falling planet.
    Initial semi-major axis as indicated. 
    Initial eccentricity: e=0.10. 
    Initial rotation periods: 2, 5, and 12.8 days.}
  \label{figure:corot29_sketch}
\end{figure}

To reduce the number of free parameters, we assumed that the
planet was a twin of CoRoT-29b in a low-eccentricity orbit. 
Figure~\ref{figure:corot29_sketch} shows the evolution of the orbit of
such a planet and the rotation of the star under the joint action of
the magnetic braking of the star and the tides on the star caused by
the planet. 
Regardless of the initial rotational rate of the star, all simulations
show that the stellar rotation quickly increases to 15 to 20 days. 
At the same time, the semi-major axis of the planet is decreasing. 
At some point the tidal torque of the planet on the star becomes
important and this accelerates the stellar rotation rate. 
By the time the planet has collided with the star, the rotational
period has been reduced to a few days. 
After the collision, the stellar rotation rate brakes to its present
value. 

The models used in the evolution simulations are  magnetic braking
as determined by \citet{bouvier1997} for stars with masses between
0.5 and 1.1 \Msun\,and a creep tide \citep{ferrazmello2013} due to a 
relaxation factor $\gamma=10s^{-1}$. 
This choice corresponds to the mid of the tidal dissipation range
determined from the statistical study of stars hosting hot Jupiters by
\citet{hansen2010} (3-25 $s^{-1}$; see \citealt{ferrazmello2013}). 
The great remaining difficulty is to discover how the existing planet
CoRoT-29b could exist in same time as its falling sibling. 
The proximity of both orbits is prone to creating great instability;
maybe the high inclination of CoRoT-29b is a consequence of this
instability. 

%
\section{Summary}
\label{section:summary}

We have discovered and characterized two new planetary systems:
CoRoT-28b and 29b.  

CoRoT-28b belongs to the small population of hot-Jupiters orbiting
evolved stars, which is interesting from the point of view of stellar
and planetary evolution. 
It has a mass between that of Saturn and Jupiter and does not seem to
be inflated \citep[see the discussion in][]{enoch2012}. 
There are a couple of open questions about the characterization of
CoRoT-28, in particular regarding its rotation. 
We have measured a higher than expected \vsini\,and its lithium
content is higher than expected for the star’s evolutionary status. 
These open questions might reveal important information about the past 
of the system. 
For example, there is the question of whether there was a former
planet that fell into the star and thus accelerated the stellar
rotation and enriched its atmosphere with heavy elements. 
Unfortunately, we cannot confirm this hypothesis because of the lack
of observational evidence. 
We  used the method described in \citet{parviainen2013} to
search for planet occultations (secondary eclipses) in the light 
curves, but we did not find any significant signal. 
For CoRoT-28b we obtain a maximum star-planet flux ratio of 9.5\%
(99th percentile of the marginal posterior). 
This upper limit is not useful for constraining the planet's albedo
or brightness temperature in any useful way, since the highest
physically feasible value for $f$ is $\sim 2\%$. 

CoRoT-29b is a very interesting system. 
The strong asymmetry of the transit light curve is only comparable to
transits of planets orbiting very young, fast rotating stars, but the
star is neither young nor rotating fast. 
The extreme value of the deformation, compared to stars of its mass
range, is not understood and deserves more attention.
We have not found at this point a satisfactory answer for the origin
of the asymmetry in the transit shape, but we encourage the
study of these system with other instruments (CHEOPS,
\citealt{broeg2013}, PLATO 2.0, \citealt{rauer2014}, ground-based
surveys) to confirm or disprove \corot\,observations and solve
the puzzle. 

%
\begin{acknowledgements}

We would like to thank the anonymous referee for comments, which led to
substantial improvements of this paper.
We are thankful for fruitful discussions to J.-P.~Zahn (LUTh,
Observatoire de Paris) and T.~Borkovits (Baja Astronomical
Observatory, Hungary).
The team at LAM acknowledges support by CNES grants 98761 (SCCB),
251091 (JMA), and 426808 (CD). RFD was supported by CNES via its
postdoctoral fellowship program. 
AS acknowledges  support from the European Research
Council/European Community under the FP7 through Starting Grant
agreement number 239953. 
AS is supported by the European Union under a Marie Curie
Intra-European Fellowship for Career Development with reference
FP7-PEOPLE-2013-IEF, number 627202. 
SCCB thanks CNES for the grant 98761.
M.A. was supported by DLR (Deutsches Zentrum f{\"u}r Luft- und
Raumfahrt) under the project 50 OW 0204.
HD and PK acknowledge support by grant AYA2012-39346-C02-02 of the
Spanish Secretary of State for R\&D\&i (MICINN).
RA acknowledges the Spanish Ministry of Economy and Competitiveness
(MINECO) for the financial support under the Ram{\'o}n y Cajal program
RYC-2010-06519, and by grant ESP2013-48391-C4-2-R. 
This research made use of data acquired with the IAC80 telescope,
operated at Teide Observatory on the island of Tenerife by the
Instituto de Astrof{\'i}sica de Canarias. 
Some of the data presented were acquired with the Nordic Optical
Telescope, operated by the Nordic Optical Telescope Scientific
Association at the Observatorio del Roque de los Muchachos, La Palma,
Spain, of the Instituto de Astrof{\'i}sica de Canarias, under OPTICON 
programme 2012A033 and CAT programme 91-NOT7/12A. 
We would like to thank the workshops and the night assistants at the
observatory in Tautenburg, Germany.
We are thankful to A.~Almazan for the observations taken for this
paper. 
This research has made use of the ExoDat database, operated at
LAM-OAMP, Marseille, France, on behalf of the CoRoT/Exoplanet
program.
This publication makes use of data products from the Two Micron All
Sky Survey, which is a joint project of the University of
Massachusetts and the Infrared Processing and Analysis
Center/California Institute of Technology, funded by the National
Aeronautics and Space Administration and the National Science
Foundation.
Funding for the Sloan Digital Sky Survey (SDSS) has been provided by
the Alfred P. Sloan Foundation, the Participating Institutions, the
National Aeronautics and Space Administration, the National Science
Foundation, the U.S. Department of Energy, the Japanese
Monbukagakusho, and the Max Planck Society. The SDSS Web site is
http://www.sdss.org/.
This research has made use of NASA's Astrophysics Data System.

\end{acknowledgements}

%
\bibliographystyle{aa}
\bibliography{bibl}

\appendix

\section{Radial velocity data}

\begin{table} 
\caption{CoRoT-28 SOPHIE radial velocities, their errors, and bisector spans.}
\label{table:rv_corot-28_SOPHIE}   
\centering
\begin{tabular}{l*{3}{c}}
\hline
BJD & RV     & Error  & Bis    \\
    & [\kms] & [\kms] & [\kms] \\  
\hline
2\,455\,801.34001   &   -76.751    &    0.023     &     -0.055   \\ 
2\,455\,804.41446   &   -76.849    &    0.013     &     -0.051   \\ 
2\,455\,810.35554   &   -76.782    &    0.021     &      0.012   \\   
2\,455\,811.32492   &   -76.753    &    0.013     &     -0.019   \\ 
2\,455\,832.30855   &   -76.723    &    0.014     &     -0.016   \\ 
2\,456\,063.56528   &   -76.771    &    0.029     &      0.054   \\   
2\,456\,064.54259   &   -76.820    &    0.015     &     -0.086   \\ 
2\,456\,071.54853   &   -76.714    &    0.017     &     -0.049   \\ 
2\,456\,072.57780   &   -76.718    &    0.010     &     -0.035   \\ 
2\,456\,100.49337   &   -76.782    &    0.012     &     -0.128   \\ 
2\,456\,103.48011   &   -76.716    &    0.010     &      0.006   \\   
2\,456\,121.38285   &   -76.810    &    0.024     &     -0.142   \\ 
2\,456\,123.47526   &   -76.671    &    0.018     &     -0.048   \\ 
2\,456\,124.46350   &   -76.593    &    0.019     &      0.016   \\   
2\,456\,125.42640   &   -76.671    &    0.013     &     -0.056   \\ 
2\,456\,132.46267   &   -76.844    &    0.018     &      0.010   \\   
2\,456\,133.40017   &   -76.778    &    0.011     &     -0.000   \\ 
2\,456\,134.43837   &   -76.720    &    0.012     &      0.023   \\   
2\,456\,135.41925   &   -76.709    &    0.018     &     -0.005   \\ 
2\,456\,149.44334   &   -76.731    &    0.017     &     -0.037   \\ 
2\,456\,152.42582   &   -76.797    &    0.016     &     -0.083   \\ 
2\,456\,153.41986   &   -76.796    &    0.010     &     -0.089   \\ 
2\,456\,154.34156   &   -76.750    &    0.022     &     -0.037   \\ 
2\,456\,156.34663   &   -76.732    &    0.016     &     -0.058   \\ 
2\,456\,157.37367   &   -76.765    &    0.013     &     -0.031   \\ 
\hline
\end{tabular}
\end{table}

\begin{table} 
\caption{CoRoT-28 HARPS radial velocities, their errors, and bisector spans.}
\label{table:rv_corot-28_HARPS}   
\centering
\begin{tabular}{l*{3}{c}}
\hline
BJD & RV     & Error  & Bis    \\
    & [\kms] & [\kms] & [\kms] \\  
\hline
2\,456\,116.75525   &   -76.7574    &    0.0070     &     -0.018   \\ 
2\,456\,118.75516   &   -76.6534    &    0.0076     &      0.013   \\   
2\,456\,152.54246   &   -76.775     &    0.013      &     -0.034   \\ 
2\,456\,158.59125   &   -76.7564    &    0.0085     &     -0.008   \\ 
2\,456\,159.50759   &   -76.7094    &    0.0088     &     -0.005   \\ 
2\,456\,161.49239   &   -76.6614    &    0.0083     &     -0.057   \\ 
2\,456\,514.54510   &   -76.6742    &    0.0085     &     -0.028   \\ 
\hline
\end{tabular}
\end{table}

\begin{table}
\caption{CoRoT-28 FIES radial velocities, their errors, and bisector spans.}
\label{table:rv_corot-28_FIES}
\centering
\begin{tabular}{l*{3}{c}}
\hline
BJD & RV     & Error  &  Bis    \\
    & [\kms] & [\kms] &  [\kms] \\
\hline
2\,456\,103.51866 &  -76.800  &   0.018  &  0.004 \\
2\,456\,104.55555 &  -76.862  &   0.022  &  0.001 \\
2\,456\,105.56260 &  -76.914  &   0.017  & -0.033 \\
2\,456\,107.54147 &  -76.844  &   0.020  & -0.032 \\
2\,456\,117.47300 &  -76.893  &   0.011  & -0.008 \\
2\,456\,118.42105 &  -76.810  &   0.017  & -0.012 \\
2\,456\,119.50754 &  -76.831  &   0.015  & -0.009 \\
2\,456\,120.55636 &  -76.888  &   0.015  & -0.034 \\
2\,456\,121.42907 &  -76.897  &   0.021  & -0.015 \\
2\,456\,122.40610 &  -76.876  &   0.021  & -0.020 \\
\hline
\end{tabular}
\end{table}

\begin{table} 
\caption{CoRoT-29 HARPS radial velocities, their errors, and bisector spans.}
\label{table:rv_corot29_HARPS}   
\centering
\begin{tabular}{l*{3}{c}}
\hline
BJD & RV     & Error  & Bis    \\
    & [\kms] & [\kms] & [\kms] \\  
\hline
2456097.73381   &   -64.131    &    0.052     &    -0.014   \\ 
2456098.70664   &   -64.371    &    0.046     &     0.022   \\   
2456099.63447   &   -64.199    &    0.103     &     0.030   \\   
2456101.81698   &   -64.303    &    0.035     &    -0.064   \\ 
2456102.80688   &   -64.097    &    0.056     &    -0.177   \\ 
2456115.66034   &   -64.302    &    0.077     &     0.105   \\   
2456116.71144   &   -64.186    &    0.026     &    -0.076   \\ 
2456117.67788   &   -64.083    &    0.021     &    -0.041   \\ 
2456119.61671   &   -64.151    &    0.051     &    -0.091   \\ 
2456149.61905   &   -64.255    &    0.080     &    -0.130   \\ 
2456150.63533   &   -64.213    &    0.037     &    -0.095   \\ 
2456151.63052   &   -64.105    &    0.045     &     0.059   \\   
2456154.57617   &   -64.053    &    0.038     &    -0.082   \\ 
2456159.55259   &   -64.096    &    0.056     &     0.008   \\   
2456161.53567   &   -64.274    &    0.037     &    -0.087   \\ 
2456455.87546   &   -64.189    &    0.102     &    -0.230   \\ 
2456460.77835   &   -64.325    &    0.099     &    -0.130   \\ 
2456508.68337   &   -64.239    &    0.041     &     0.017   \\   
2456511.54785   &   -64.228    &    0.031     &    -0.130   \\ 
2456514.68026   &   -64.315    &    0.048     &    -0.145   \\ 
\hline
\end{tabular}
\end{table}

\section{Flow chart for gravity darkening modelling}

\begin{figure*}
  \centering
  \includegraphics[%
  width=0.9\linewidth,%
  height=0.5\textheight,%
  keepaspectratio]{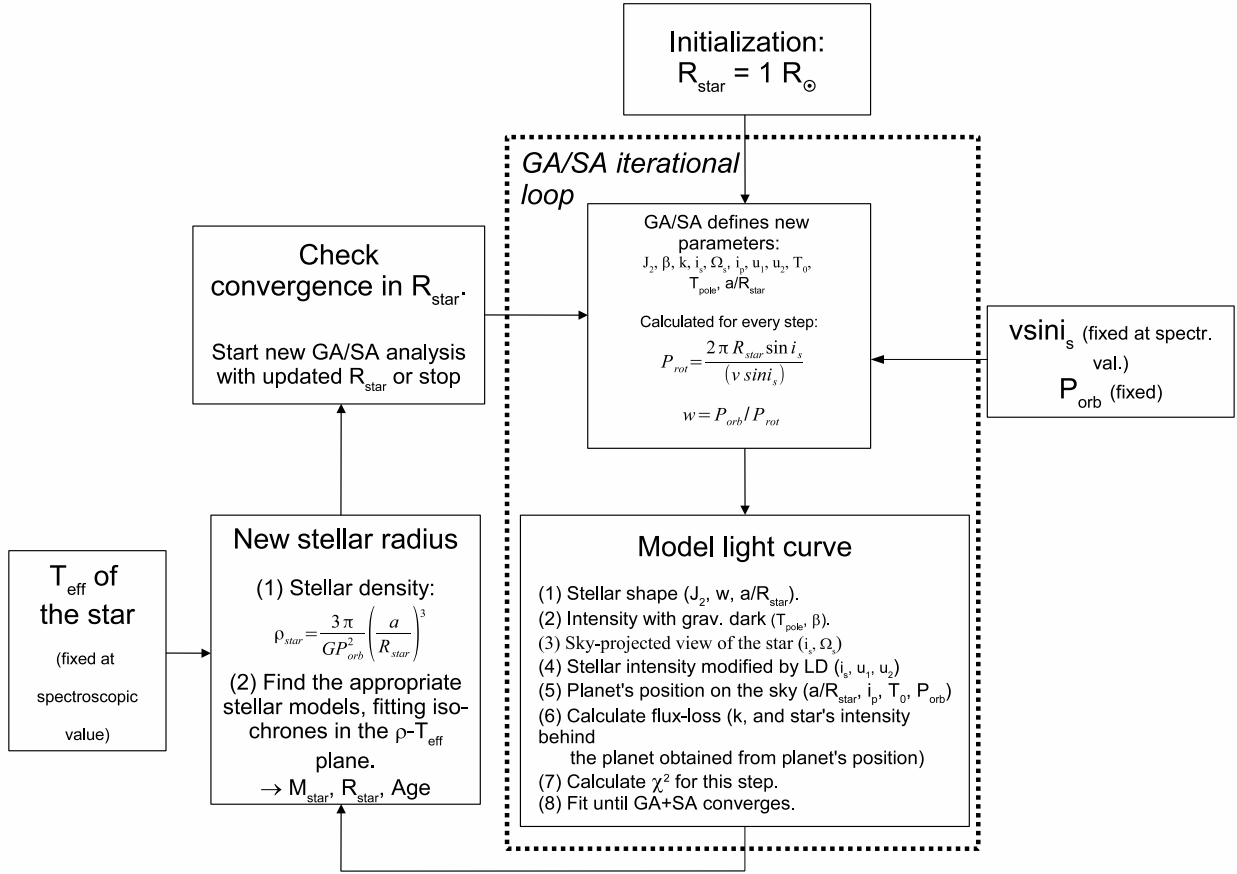}
  \caption{Flow chart for the gravity darkening modelling process of
    CoRoT-29b.} 
  \label{figure:gravity_darkening_flow_chart}. 
\end{figure*}

\end{document}